\documentclass[twocolumn,showpacs,preprintnumbers,amsmath,amssymb,nofootinbib]{revtex4}

\usepackage{graphicx}
\usepackage{dcolumn}
\usepackage{bm}

\begin{document}


\title{Exact conformal field theories from  mutually T-dualizable $\sigma$-models}

\author{Ali Eghbali}
 \altaffiliation[eghbali978@gmail.com]{}
\affiliation{%
Department of Physics, Faculty of Basic Sciences, \\
Azarbaijan Shahid Madani University, 53714-161, Tabriz, Iran
}%

\date{\today}

\begin{abstract}
Exact conformal field theories (CFTs) are obtained by using the approach of
Poisson-Lie (PL) T-duality in the presence of spectators.
We explicitly construct some
non-Abelian T-dual $\sigma$-models (here as the PL T-duality on a semi-Abelian double) on $2+2$-dimensional
target manifolds $M \approx O \times \bf G$ and
${\tilde M} \approx O \times {\bf {\tilde G}}$, where $\bf G$ and ${\bf {\tilde G}}$ as two-dimensional
real non-Abelian and Abelian Lie groups act freely on $M$ and $\tilde M$, respectively, while $O$ is the orbit of $\bf G$ in $M$.
The findings of our study show that the original models are  equivalent to
Wess-Zumino-Witten (WZW) models based on the Heisenberg  $(H_4)$
and $GL(2,\mathbb{R})$ Lie groups. In this way,  some new T-dual  backgrounds for
these  WZW models are obtained. For one of the duals of   the $H_4$  WZW model, we show that the model is self-dual.
In the case of the $GL(2,\mathbb{R})$ WZW model it is observed that
the duality transformation  changes the asymptotic behavior of solutions from $AdS_{3} \times \mathbb{R}$ to flat space.
Then, the structure and  asymptotic nature of the dual
spacetime  of this model including the horizon and singularity are determined.
We furthermore get the non-critical
Bianchi type III string cosmological model with a non-vanishing field strength from T-dualizable $\sigma$-models
and show that this model describes an exact CFT (equivalent to the $GL(2,\mathbb{R})$ WZW model).
After that, the conformal invariance of T-dual models up to two-loop order (first order in $\alpha'$) is discussed.
\end{abstract}

\pacs{11.25.Tq, 11.25.Hf, 11.25.Pm}
\maketitle

\section{Introduction}

The duality symmetries play an
important role in string theory. On the one hand, they are specific to string
theory and their study has led to important insights in understanding
the spacetime geometry  from the string point of view. A very
important symmetry of string theory or more generally,
two-dimensional sigma models, is the T-duality \cite{Buscher1}. A study of the T-duality in string theory has led to the discovery
of PL T-duality.
Klim\v{c}ik and \v{S}evera in their seminal work \cite{Klim1}
proposed a generalization of T-duality, or the so-called  PL T-duality, which allows the duality to be performed on a target space without isometries.
In  Klim\v{c}ik and  \v{S}evera's formalism,  PL T-dual
sigma models are defined by PL group manifolds which constitute a Drinfeld double \cite{Drinfeld}.
The classification of low-dimensional Drinfeld doubles \cite{{JR},{Hlavaty1}} has become a convenient laboratory for
investigation of the PL T-duality.

On the other hand, the duality symmetries in  WZW models  have received
considerable attention because of the preservation of  the conformal symmetry
under the Abelian duality \cite{Verlinde}.
This duality has been investigated in the WZW models \cite{Kiritsis}. Furthermore, for the case of non-Abelian duality \cite{Ossa},
it has been shown that the conformal
symmetry is preserved when the trace of the adjoint representation of the isometry group is zero \cite{Givoen}.

The WZW model is a well-known construction for obtaining a CFT which describes string propagation on a Lie group.
For instance, the natural metric on the Lie group $SL(2,\mathbb{R})$ is precisely the three-dimensional anti-de Sitter
metric. Hence, the WZW model based on Lie group $SL(2,\mathbb{R})$ can be considered as an exact CFT describing
string propagation on anti-de Sitter space \cite{Balog1}.
Up to now, only few examples of PL  symmetric $\sigma$-models  have been treated at the
quantum level \cite{{Alekseev1},{Lledo}}.
Furthermore, PL symmetry in the WZW models based on the Lie supergroups have recently been studied
in Refs. \cite{{ER7},{ER8}}.
We also refer the reader to the literatures given in Ref. \cite{Sfetsos3}.
In Ref. \cite{Alekseev1} it has been shown that the duality relates the $SL(2 , \mathbb{R})$ WZW model to a constrained
$\sigma$-model defined by the $SL(2,\mathbb{R})$ group space.
We have shown that \cite{eghbali11} the PL T-duality relates the
$H_4$ WZW model to a $\sigma$-model defined on the dual Lie group ${ A}_2 \oplus 2{ A}_1$.
We have also stressed that the dual model is conformally invariant up to two-loop order.
Furthermore, we have recently shown that \cite{EMR13} the PL T-duality relates the
$SL(2 , \mathbb{R})$ WZW model to a $\sigma$-model defined on $2+1$-dimensional
manifold ${M} \approx O \times {\bf G}$ in which ${\bf G}$ is
two-dimensional real non-Abelian Lie group $A_2$, and $O$ as a one-dimensional space is
the orbit of ${\bf G}$ in ${ M}$. Accordingly, we have obtained
a dual model for the $SL(2 , \mathbb{R})$ WZW model yielding
a new three-dimensional charged black string which is stationary and asymptotically flat.

The main purpose of this paper is to construct some new non-Abelian T-dual backgrounds for the $H_4$ and $GL(2,\mathbb{R})$  WZW models via
PL T-duality approach in the presence of spectators.
The original models as exact CFTs (the $H_4$ and $GL(2,\mathbb{R})$  WZW models)
are constructed on $2+2$-dimensional target manifold $M \approx O \times \bf G$ with ${\bf G } = A_2$
and dual models on manifold ${\tilde M} \approx O \times {\bf {\tilde G}}$ with ${\tilde {\bf G }} = 2A_1$,
whereas, in \cite{eghbali11} T-dual $\sigma$-models were only  constructed on Lie groups in the absence of spectators.
In the present work, two dual models for the $H_4$ WZW  are obtained for one  of which
we show that the dual model is indeed identical to the same  $H_4$  WZW model.
Moreover, we get one dual model
for the $GL(2,\mathbb{R})$ WZW for which
the structure and  asymptotic nature of the
spacetime  including the horizon and singularity are determined.
We also obtain the non-critical
Bianchi type III string cosmological model with a non-vanishing field strength from a T-dualizable $\sigma$-model
on $3+1$-dimensional target manifold $M \approx O \times \bf G$, in which
$\bf G$ represents three-dimensional decomposable Lie group $A_2 \oplus A_1$,
and then we show that this model describes an exact CFT.
Finally, we discuss the conformal invariance conditions of the T-dual models up to the first order in $\alpha'$
to  introduce new solutions for two-loop  $B$-function equations of the  $\sigma$-model with a non-vanishing field strength
$H$ and the dilaton field in both cases of the
absence and presence of a cosmological constant $\Lambda$.

This paper is organized as follows. In Sec. \ref{Sec.II},  we present a basic review of the
PL T-dual $\sigma$-models construction in the presence of spectator fields.
In Sec. \ref{Sec.III}, we get the $H_4$ and $GL(2,\mathbb{R})$ WZW models from
T-dualizable $\sigma$-models constructed on $2+2$-dimensional target manifolds $M \approx O \times \bf G$
and  ${\tilde M} \approx O \times {\bf {\tilde G}}$. In addition, the dual backgrounds for these WZW models
together with the structure and  asymptotic nature of the dual spacetime of the
$GL(2,\mathbb{R})$ WZW
including the horizon and singularity are studied. Finally, the non-Abelian T-dualization of
the non-critical Bianchi type III string cosmology solution is discussed at  the end of Sec. \ref{Sec.III}.
In Sec. \ref{Sec.IV}, we investigate the
conformal invariance conditions for T-dual models up to two-loop order.
Some concluding remarks are given in Sec. \ref{Sec.V}.

\section{\label{Sec.II}Construction of PL T-dual $\sigma$-models with spectators}

We begin this section by reviewing the construction of PL T-dual $\sigma$-models in the presence of spectator fields. First of all, for the
description of  PL T-duality we need to introduce the Drinfeld double group $\bf D$ \cite{Drinfeld}, which by definition has a pair of
maximally isotropic subgroups $\bf G$ and ${\tilde {\bf G}}$ corresponding to the subalgebras ${\cal G}$ and ${\tilde {\cal G}}$, respectively.
The generators of  ${\cal G}$ and ${\cal \tilde G}$ are denoted, respectively, $T_a$ and ${\tilde T}^a$, $a=1,\cdots,dim~{\bf G}$.
One says that the Lie algebras ${\cal G}$ and ${\tilde {\cal G}}$ are compatible if the brackets
\begin{eqnarray}
[T_a , T_b] &=& {f^c}_{ab} ~T_c,~~~
[{\tilde T}^a , {\tilde T}^b] = {{\tilde f}^{ab}}_{\; \; \: c} ~{\tilde T}^c,\nonumber\\
{[T_a , {\tilde T}^b]} &=& {{{\tilde f}^{bc}}_{\; \; \; \:a} {T}_c + {f^b}_{ca} ~{\tilde T}^c},
\end{eqnarray}
define a Lie algebra structure on the direct sum vector space ${\cal D} = {\cal G} \oplus {\tilde {\cal G}}$.
In this case, we say that the Lie algebra ${\cal D}$ is the Drinfeld double of ${\cal G}$ or, equivalently, of ${\tilde {\cal G}}$.
Thus, the group $\bf D$ is called the Drinfeld double of $\bf G$ (or ${\tilde {\bf G}}$). We also note that
the Drinfeld double ${\cal D}$ is equipped with an invariant inner product
$<.~ ,~ .>$ with the following properties
\begin{eqnarray}
<T_a , {\tilde T}^b> &=& {\delta _a}^{~b},\nonumber\\
 <T_a , T_b> &=& <{\tilde T}^a, {\tilde T}^b> ~ =~ 0.
\end{eqnarray}
In what follows we shall investigate PL T-duality transformations in the presence of spectators \cite{{Klim1},{Sfetsos1}}
of a non-linear $\sigma$-model with the following  action for a bosonic string,
propagating in a $d$-dimensional  spacetime, with the metric ${G}_{_{\mu\nu}}$,
the antisymmetric tensor field ${B}_{_{\mu\nu}}$ and the dilaton field $\phi$
\begin{eqnarray}\label{a.3}
S &=& \frac{1}{2 \pi \alpha'}\int_{{\Sigma}}\!d\tau  d\sigma \sqrt{-h} \Big[\frac{1}{2}\big(h^{\alpha \beta}G_{_{\mu\nu}}
+\epsilon^{\alpha \beta} B_{_{\mu\nu}}\big)\partial_{\alpha}x{^{^\mu}}
\partial_{\beta}x^{^{\nu}} \nonumber\\
&&~~~~~~~~~~~~~~~~~~~+ \frac{1}{4} \alpha'  \phi ~ R^{^{(h)}}\Big],
\end{eqnarray}
where $h_{\alpha \beta}$ is the worldsheet metric with $R^{^{(h)}}$ the corresponding  worldsheet curvature scalar and
$h=\det h_{\alpha \beta}$. The indices $\alpha, \beta$ run over $(\tau , \sigma)$, and $\epsilon^{\alpha \beta}$ is
an antisymmetric tensor on the worldsheet ${\Sigma}$.
The dimensionful coupling constant $\alpha'$ turns out to be the inverse string tension.
The functions $x^{^{\mu}}:{\Sigma} \longrightarrow \mathbb{R}$, (${\mu} =1,...,dim~M$)
are obtained by the composition $x^{^{\mu}}={X^{^{\mu}}} \circ x$ of a map
$x:{\Sigma} \longrightarrow M$ and components of a coordinate map ${X}$ on a chart of $M$.
Here, and in the following, we use the standard light-cone variables on the worldsheet, $\sigma^{\pm} = \tau \pm \sigma$.

Let us now consider a $d$-dimensional manifold $M$ and some coordinates $x^{^{\mu}} = (x^{i} , y^\alpha)$ on it,
where $x^i (i = 1,\cdots,dim~{\bf G})$
are the coordinates of Lie group $\bf G$ acting freely from right on $M$. $y^\alpha~(\alpha = 1,\cdots,d-dim~{\bf G})$ are
the  coordinates labeling the orbit $O$ of $\bf G$ in the target space $M$.
We note that  the coordinates $y^\alpha$ do not participate in the PL T-duality transformations
and are therefore called spectators \cite{Sfetsos1}.
Take a linear (idempotent) map  ${\cal K}$ from the space
$T_y^{\ast} M \oplus T_y M \oplus {\cal D}$ into itself. It has two eigenspaces ${\mathbb{R}}_{\pm}(y^{\alpha})$ with eigenvalues $\pm 1$.
They are perpendicular to each other according to the bilinear form on $T_y^{\ast} M \oplus T_y M \oplus {\cal D}$.
These eigenspaces may be considered as the graph of a non-degenerate linear map
$E^{\pm}(y)$: $ T_y M \oplus  {{\cal G}} \longrightarrow T_y^{\ast}
M \oplus {\tilde {\cal G}}$, such that by translating this graph to the point ${g} \in {\bf G}$ we have
\begin{equation}\label{a.4}
g^{-1} {\mathbb{R}}_{\pm}(y^{\alpha}) g\;=\; Span\{X_{_A} \pm E^{\pm}_{_{AB}}(g,y^{\alpha}) {\tilde X}^{^B}\},
\end{equation}
where $X_A =(T_a ,  {\partial}_{\alpha})$
and ${\tilde X}^A =({\tilde T}^a , {dy}^{\alpha})$ are the basis of the  spaces $T_y M \oplus
{{\cal G}}$ and $T_y^{\ast} M \oplus {\tilde  {{\cal G}}}$, respectively. In order to determine the $d \times d$ matrix $E^{\pm}_{AB}(g,y^{\alpha})$ we  write
the spaces ${\mathbb{R}}_{\pm}(y^{\alpha})$ as follows:
\begin{eqnarray}\label{a.5}
g^{-1} {\mathbb{R}}_{\pm}(y^{\alpha}) g= Span\{g^{-1} {X_{_A}} g
\pm E^{\pm}_{_{AB}}(e,y^{\alpha}) g^{-1} {\tilde X}^{^B} g \},~
\end{eqnarray}
in which the  matrix $E^{\pm}_{AB}(e,y^{\alpha})$ is defined as
\begin{eqnarray}\label{a.6}
E^{\pm}_{_{AB}}(e,y^{\alpha})\;=\;\left( \begin{array}{cc}
                     E^{\pm}_{0\;ab}\tiny(e,y^{\alpha}) & F^{\pm^{(1)}}_{a \beta}(e,y^{\alpha})\\

                     F^{\pm^{(2)}}_{\alpha b}(e,y^{\alpha}) & F_{\alpha \beta}(y^{\alpha})
                      \end{array} \right).
\end{eqnarray}
Here, submatrices $E^{\pm}_{0\;ab}(e,y^{\alpha})$,
$F^{\pm^{(1)}}_{a \beta}(e,y^{\alpha})$ and $ F^{\pm^{(2)}}_{\alpha b}(e,y^{\alpha}) $ are functions of the variables $y^{\alpha}$ and $e$, where
$e$ is the unit element of $\bf G$.  $F_{\alpha \beta}(y^{\alpha})$ is also a function of $y^{\alpha}$ only.
Here, and in the following, the minus sign stands for transpose, namely,
$E^{+}_{0\;ab}=E^{-}_{0\;ba}$, $F^{+^{(1)}}_{a \beta}=F^{-^{(2)}}_{\beta a}$ and $F^{+^{(2)}}_{\alpha b}=F^{-^{(1)}}_{b \alpha}$.

It is convenient to define matrices $a(g)$, $b(g)$ and the Poisson bracket $\Pi(g)$ in the following way
\begin{eqnarray}
g^{-1} T_{{_a}}~ g &=& a_{_{a}}^{^{~b}}(g) ~ T_{{_b}},\nonumber\\
g^{-1} {\tilde T}^{{^a}} g &=&
b^{^{ab}}(g)~ T_{{_b}}+(a^{-1})_{_{b}}^{^{~a}}(g)~{\tilde T}^{{^b}},\label{a.7}\\
\Pi^{^{ab}}(g) &=& b^{^{ac}}(g)~ (a^{-1})_{_{c}}^{^{~b}}(g).\label{a.8}
\end{eqnarray}
Thus, using \eqref{a.4} and \eqref{a.5} together with \eqref{a.7} one gets
\begin{eqnarray}\label{a.9}
E^{\pm}_{AB}(g,y^{\alpha})&=&{\big(A(g)\pm E^{\pm}(e,y^{\alpha}) B(g)\big)^{-1}}_{_{\hspace{-3mm}A}}^{^{~C}}\nonumber\\
&&~~\times E^{\pm}_{_{CD}}(e,y^{\alpha})~
(A^{-1})_{_{B}}^{^{~D}}(g),
\end{eqnarray}
where \footnote{Here Id means the identity matrix.}
\begin{eqnarray}\label{a.10}
A(g)\;=\;\left( \begin{array}{cc}
                     a(g) & 0\\

                     0 & Id
                      \end{array} \right),~~
                      B(g)\;=\;\left( \begin{array}{cc}
                     b(g) & 0\\

                     0 & 0
                      \end{array} \right).
\end{eqnarray}
We also define
\begin{eqnarray}\label{a.11}
\mathbb{F}^{\pm}_{_{AB}}(g,y^{\alpha})\;=\;A_{_{A}}^{^{~C}}(g)~E^{\pm}_{_{CD}}(g,y^{\alpha})~A_{_{B}}^{^{~D}}(g).
\end{eqnarray}
Considering matrix $\mathbb{F}^{\pm}_{AB}(g,y^{\alpha})$ in the form
\begin{eqnarray}\label{a.12}
\mathbb{F}^{\pm}_{_{AB}}(g,y^{\alpha})\;=\;\left( \begin{array}{cc}
                     {\mathbb{E}}^{\pm}_{ab}\tiny(g,y^{\alpha}) & \Phi^{\pm^{(1)}}_{a \beta}(g,y^{\alpha})\\

                     \Phi^{\pm^{(2)}}_{\alpha b}(g,y^{\alpha}) & \Phi_{_{\alpha \beta}}(y^{\alpha})
                      \end{array} \right),
\end{eqnarray}
and then using \eqref{a.6}, \eqref{a.9}, \eqref{a.10} and \eqref{a.11} one can obtain the backgrounds appearing
in the action of original $\sigma$-model. They are given in matrix notation by
{\small \begin{eqnarray}
{\mathbb{E}}^{\pm}\tiny(g,y^{\alpha})&=&\Big(E^{\pm^{-1}}_{0}\tiny(e,y^{\alpha})\pm \Pi(g)\Big)^{-1},\label{a.13}\\
\Phi^{\pm^{(1)}}(g,y^{\alpha})&=& {\mathbb{E}}^{\pm}\tiny(g,y^{\alpha})(E^{\pm}_{0})^{-1}\tiny(e,y^{\alpha})F^{\pm^{(1)}}(e , y^{\alpha}),\label{a.14}\\
\Phi^{\pm^{(2)}}(g,y^{\alpha})&=& F^{\pm^{(2)}}(e , y^{\alpha}) (E^{\pm}_{0})^{-1}\tiny(e,y^{\alpha}){\mathbb{E}}^{\pm}\tiny(g,y^{\alpha}),\label{a.15}\\
\Phi(g,y^{\alpha})&=& F(y^{\alpha})-F^{+^{(2)}}(e , y^{\alpha})\Pi(g){\mathbb{E}}^{+}\tiny(g,y^{\alpha})\nonumber\\
&&~~~~~~\times (E^{+}_{0})^{-1}\tiny(e,y^{\alpha})~F^{+^{(1)}}(e , y^{\alpha}).\label{a.16}
\end{eqnarray}}
Let us now introduce the elements $V_{\pm}$ of subspaces ${\mathbb{R}}_{\pm}(y^{\alpha})$ as
\begin{eqnarray}
V_{\pm}~:=~\partial_{\pm} y^{\alpha}~\frac{\partial}{\partial y^{\alpha}} \mp p_{\alpha}^{(\mp)} dy^{\alpha} + \partial_{\pm}l l^{-1},\label{a.17}
\end{eqnarray}
where $p_{\alpha}^{(\mp)} \in T_y^{\ast} M$ and $l\in \bf D$. Inserting the decomposition $l = g {\tilde h} ~(g\in {\bf G}, {\tilde h}\in {\tilde {\bf G}})$
\cite{Alekseev1} into \eqref{a.17} we get
\begin{eqnarray}
V_{\pm}~:=~R_{\pm}^{^A} X_{_A} + {\tilde R}_{\pm_{A}}  \big(B^{^{BA}} (g)~ X_{_B}+A_{_{B}}^{~~^{A}}(g){\tilde X}^{^B}\big),~~~\label{a.18}
\end{eqnarray}
where $R_{\pm}^{^A}$ and ${\tilde R}_{\pm_{A}}$ are the elements of the respective spaces $T_y M \oplus
{{\cal G}}$ and $T_y^{\ast} M \oplus {\tilde  {{\cal G}}}$, and are given by
\begin{eqnarray}
R_{\pm}^{^A} &=& \big(R_{\pm}^{^a}~,~\partial_{\pm} y^{\alpha}\big) = \big((\partial_{\pm} g g^{-1})^a~,~\partial_{\pm} y^{\alpha}\big),\\\label{a.19}
{\tilde R}_{\pm_{A}}&=& \big((\partial_{\pm} {\tilde h} {\tilde h}^{-1})_a~,~\mp p_{\alpha}^{(\mp)}\big).\label{a.20}
\end{eqnarray}
Thus, by using the equations of motion
\begin{eqnarray}
\Big<V_{\pm} ~,~ \big( X_{_A} \mp E^{\mp}_{_{AB}}(g,y^{\alpha})~ {\tilde X}^{^B}\big)\Big> ~=~0, \label{a.21}
\end{eqnarray}
we obtain
\begin{eqnarray}
{\tilde R}_{\pm_{A}}~=~\pm R_{\pm}^{^B}~\mathbb{F}^{\pm}_{_{BC}}(g,y^{\alpha}) ~ (A^{-1})_{_{A}}^{^{~C}}(g). \label{a.22}
\end{eqnarray}
The equation \eqref{a.22} can be written in terms of components. They then take the following forms
\begin{eqnarray}
(\partial_{+} {\tilde h} {\tilde h}^{-1})_a &=& (a^{-1})_{a}^{{~c}}(g) \Big[R_+^b~{\mathbb{E}}^{+}_{bc}\tiny(g,y^{\alpha})\nonumber\\
&&~~~~~~~~~~~+\partial_+ y^\alpha~ \Phi^{+^{(2)}}_{\alpha c}(g,y^{\alpha})\Big],\\\label{a.23}
(\partial_{-} {\tilde h} {\tilde h}^{-1})_a &=& -(a^{-1})_{a}^{{~c}}(g) \Big[{\mathbb{E}}^{+}_{cb}\tiny(g,y^{\alpha})~R_-^b\nonumber\\
&&~~~~~~~~~~~+\Phi^{+^{(1)}}_{c \beta}(g,y^{\alpha})~ \partial_- y^\beta\Big],\label{a.24}
\end{eqnarray}
and
\begin{eqnarray}
p_{\alpha}^{(+)} &=& -\Big[\Phi^{+^{(2)}}_{\alpha b}(g,y^{\alpha})~R_-^b + \Phi_{_{\alpha \beta}}(g,y^{\alpha})~ \partial_- y^\beta\Big],~~\\\label{a.25}
p_{\alpha}^{(-)} &=& -\Big[R_+^a~\Phi^{+^{(1)}}_{a \alpha}(g,y^{\alpha}) + \partial_+ y^\beta~ \Phi_{_{\beta \alpha}}(g,y^{\alpha})\Big].~~\label{a.26}
\end{eqnarray}
The above results indicate that the equations \eqref{a.21} are nothing but the equations of motion
concerning  the $\sigma$-model described by the following action
\begin{eqnarray}\label{a.27}
S &=&\frac{1}{2} \int d\sigma^{+}  d\sigma^{-}  ~{\mathbb{F}_{_{AB}}^{+}}(g,y^{\alpha})~{R_+^A}\;{R_-^B},\nonumber\\
&=&\frac{1}{2} \int d\sigma^{+}  d\sigma^{-}\Big[{\mathbb{E}_{_{ab}}^{+}}(g,y^{\alpha})~
{R_+^a}\;{R_-^b}\nonumber\\
&&~~+\Phi^{+^{(1)}}_{a \beta}(g,y^{\alpha}) {R_+^a}\partial_{-} y^{\beta}+
\Phi^{+^{(2)}}_{\alpha b}(g,y^{\alpha}) \partial_{+} y^{\alpha} {R_-^b}\nonumber\\
&&~~+\Phi_{_{\alpha \beta}}(g,y^{\alpha})
\partial_{+} y^{\alpha} \partial_{-} y^{\beta}\Big].
\end{eqnarray}
As we shall see below, one can construct another $\sigma$-model (denoted as usual
with tilded symbols) which is said to be dual to \eqref{a.27}
in the sense of the PL T-duality if the Lie algebras $\cal G$ and ${\cal \tilde G}$ form a pair of maximally isotropic
subalgebras of the Lie algebra $\cal D$.
In order to get the dual $\sigma$-model one proceeds in an analogous way so that eigenspaces ${\mathbb{R}}_{\pm}(y^\alpha)$ are
considered as
\begin{eqnarray}\label{a.28}
{\mathbb{R}}_{\pm}(y^{\alpha}) \;=\; Span\{{{\tilde Y}^{^A}}
\pm {{\tilde E}^{\pm^{AB}}}({\tilde e} , y^{\alpha}) ~ { Y}_{_B}\},
\end{eqnarray}
where ${{\tilde E}}^+(y^{\alpha}):~T_y M \oplus {\tilde  {{\cal G}}} \longrightarrow T_y^{\ast} M \oplus {\cal G}$ ,
${Y}_{_B} = (T_a , d y^\alpha)$ and ${{\tilde Y}^{^A}} = ({\tilde T}^a , \partial_{\alpha})$. With a slight abuse of the notation,
comparing \eqref{a.5} and \eqref{a.28} we get the matrix form of ${{\tilde E}^{\pm}}({\tilde e} , y^{\alpha})$ as \cite{Klim1}
\begin{eqnarray}\label{a.29}
{{\tilde E}^{\pm}}({\tilde e} , y^{\alpha}) &=&\pm \big({\cal A}\pm {{E}^{\pm}}({e} , y^{\alpha})~{\cal B}\big)^{-1}\nonumber\\
&&~~~~\times \big({\cal B}\pm {{E}^{\pm}}({e} , y^{\alpha})~{\cal A}\big),
\end{eqnarray}
in which
\begin{eqnarray}\label{a.30}
{\cal A}\;=\;\left( \begin{array}{cc}
                     0 & 0\\

                     0 & Id
                      \end{array} \right),~~~~~~~~~
                    {\cal B}\;=\;\left( \begin{array}{cc}
                     Id & 0\\

                     0 & 0
                      \end{array} \right).
\end{eqnarray}
Now, using \eqref{a.28} and inserting the decomposition $l = {\tilde g} h$ into the equations \eqref{a.21},
one can get the equations of motion for $y^\alpha$ and ${\tilde x}^i$ corresponding to the following action
\begin{eqnarray}\label{a.31}
\tilde S &=& \frac{1}{2} \int d\sigma^{+}  d\sigma^{-} ~{\mathbb{{\tilde F}}^{+^{AB}}}(\tilde g , y^{\alpha})~{\tilde R}_{+_{A}}~{\tilde R}_{-_{B}}\nonumber\\
&=&\frac{1}{2} \int d\sigma^{+}  d\sigma^{-}\Big[{{\mathbb{\tilde E}}^{+^{ab}}}(\tilde g , y^{\alpha})~
{\tilde R}_{+_{a}}\;{\tilde R}_{-_{b}}\nonumber\\
&&~~+{\tilde \Phi}^{\hspace{0mm}+^{(1)^{ a}}}_{~~~~\beta}(\tilde g,y^{\alpha}) {\tilde R}_{+_{a}}\partial_{-} y^{\beta}+
{\tilde \Phi}^{\hspace{0mm}+^{(2)^{ b}}}_{\alpha}(\tilde g,y^{\alpha}) \partial_{+} y^{\alpha} ~{\tilde R}_{-_{b}}\nonumber\\
&&~~+{\tilde \Phi}_{_{\alpha \beta}}(\tilde g,y^{\alpha})
\partial_{+} y^{\alpha} \partial_{-} y^{\beta}\Big].
\end{eqnarray}
The coupling matrices of the dual $\sigma$-model are also determined in a similar fashion \cite{{Klim1},{Sfetsos1}}.
Using \eqref{a.29} one relates them to those of the original one by
\begin{eqnarray}
{\mathbb{\tilde E}}^{\pm}\tiny(\tilde g,y^{\alpha})&=&\Big(E^{\pm}_{0}\tiny(e,y^{\alpha})\pm \tilde \Pi(\tilde g)\Big)^{-1},\label{a.32}\\
{\tilde \Phi}^{\pm^{(1)}}(\tilde g,y^{\alpha})&=& \pm~ {\mathbb{\tilde E}}^{\pm}\tiny(\tilde g,y^{\alpha})~F^{\pm^{(1)}}(e , y^{\alpha}),\label{a.33}\\
{\tilde \Phi}^{\pm^{(2)}}(\tilde g,y^{\alpha})&=& \mp~ F^{\pm^{(2)}}(e , y^{\alpha}) ~{\mathbb{\tilde E}}^{\pm}\tiny(\tilde g,y^{\alpha}),\label{a.34}\\
{\tilde \Phi}(\tilde g,y^{\alpha})&=& F(y^{\alpha})-F^{+^{(2)}}(e , y^{\alpha})\nonumber\\
&&~~~~~~\times {\mathbb{\tilde E}}^{+}\tiny(\tilde g,y^{\alpha})~F^{+^{(1)}}(e , y^{\alpha}).\label{a.35}
\end{eqnarray}
The actions  \eqref{a.27} and \eqref{a.31} correspond to PL T-dual $\sigma$-models \cite{Klim1}. Notice that if the group
${\bf G}(\tilde {\bf G})$ besides having free action on $M(\tilde M)$, acts transitively on it, then the corresponding
manifold  $M(\tilde M)$ will be the same as the group ${\bf G}(\tilde {\bf G})$. In this case only the first term appears
in the actions \eqref{a.27} and \eqref{a.31}.

In the PL T-duality case, dilaton shifts in both models have been
obtained by quantum considerations based on a regularization of a functional
determinant in a path integral formulation of PL T-duality by
incorporating spectator fields \cite{Tyurin} (see, also, \cite{N.Mohammedi})
\begin{eqnarray}
\phi &=&\phi_{_{0}}(y^{\alpha})+\log (\det {\mathbb{E}}^{+}) - \log (\det {E_0^+}),\label{a.35.1}\\
{\tilde \phi} &=&\phi_{_{0}}(y^{\alpha})+\log (\det {\mathbb{\tilde E}}^{+}),\label{a.35.2}
\end{eqnarray}
where  $\phi_{_{0}}(y^{\alpha})$ is just a function of $y^{\alpha}$.

\section{\label{Sec.III}T-dualizable $\sigma$-models on $2+2$-dimensional manifolds as exact CFTs}

In this section, we explicitly construct  two pairs of PL T-dual $\sigma$-models on $2+2$-dimensional target
manifolds $M \approx O \times {\bf G}$ and ${\tilde M} \approx O \times {\tilde {\bf G}}$, where $\bf G$ and $\tilde {\bf G}$ as two-dimensional
real non-Abelian and Abelian Lie groups act freely on $M$ and $\tilde M$, respectively, while $O$ is the orbit of ${\bf G}$ in $M$ with the
spectators $y^\alpha =\{y_1, y_2\}$.
 The  Lie algebras of the Lie groups ${\bf G}$ and $\tilde {\bf G}$ are denoted by
${\cal A}_2$ and $2{\cal A}_1$, respectively.
According to Sec. \ref{Sec.II},  having  Drinfeld doubles we can construct PL T-dual $\sigma$-models on them.
The four-dimensional Lie algebra of the Drinfeld double $({\cal A}_2 , 2{\cal A}_1)$ is given by the following non-zero commutation relations:
\begin{eqnarray}\label{d.1}
[T_1 , T_2]~=~T_2,~~[T_1 ~, ~{\tilde T}^2]=-{\tilde T}^2,~~[T_2 ~, ~{\tilde T}^2]={\tilde T}^1,
\end{eqnarray}
where $\{T_1 , T_2\}$ and $\{{\tilde T}^1 , {\tilde T}^2\}$ are the basis of ${\cal A}_2$ and $2{\cal A}_1$, respectively.
Notice that  the double $({\cal A}_2 , 2{\cal A}_1)$ has non-vanishing trace in the adjoint representations.
In such a situation, there is usually a conformal anomaly at one-loop associated with non-Abelian T-duality \cite{N.Mohammedi}.
In what follows, we will also discuss the conformal anomaly appeared in the string effective Lagrangians corresponding to the T-dual models.

In order to calculate the components of right invariant one-forms
$R_{\pm}^a$ on the Lie group $A_2$ we parametrize an element of $A_2$ as
\begin{eqnarray}\label{d.2}
g~=~e^{x_1 T_1}~e^{x_2 T_2},
\end{eqnarray}
where ${ x}^i = \{x_1, x_2\}$  are the coordinates of the Lie group $A_2$. $R_{\pm}^a$'s are then derived in the following form
\begin{eqnarray}\label{d.3}
R_{\pm}^1~=~ \partial_{\pm} x_1,~~~~~~~~~~~~R_{\pm}^2~=~ e^{x_1}~\partial_{\pm} x_2.
\end{eqnarray}
Since the dual Lie group, $2A_1$, is Abelian,
 by using \eqref{a.7}, \eqref{a.8} and \eqref{d.1} it follows  that the Poisson bracket
$\Pi^{ab}(g)$ on $A_2$ vanishes. Furthermore,  for obtaining the Poisson bracket on
the dual group $2A_1$ we first parametrize the Lie group $2A_1$ with
coordinates ${\tilde x}^i = \{{\tilde x}_1 , {\tilde x}_2\}$ so that its elements
are defined as in \eqref{d.2} by replacing untilded quantities with tilded ones. Then,
using \eqref{a.7} and \eqref{a.8} for tilded quantities
together with \eqref{d.1} the Poisson bracket on  $2A_1$  is derived as follows:
\begin{eqnarray}\label{d.9}
{\tilde \Pi}_{ab}(\tilde g)\;=\;\left( \begin{array}{cc}
                     0 & -{\tilde x}_2\\

                     {\tilde x}_2 & 0
                      \end{array} \right).
\end{eqnarray}
In addition to the right invariant one-forms, to construct the $\sigma$-models \eqref{a.27} and \eqref{a.31} on manifolds $M$ and $\tilde M$
we need to determine the couplings $\mathbb{E}^{+}_{ab}\tiny(g , y^\alpha)$,
$\Phi^{+^{(1)}}_{a \beta}\tiny{(g , y^\alpha)}$, $\Phi^{+^{(2)}}_{\alpha b}\tiny{(g , y^\alpha)}$ and $\Phi_{\alpha\beta} (g,y^\alpha)$.
By convenient choices of
the background matrices $E^{+}_{0\;ab}\tiny(e , y^\alpha)$, $F^{+^{(1)}}_{a \beta}\tiny(e , y^\alpha)$, $F^{+^{(2)}}_{\alpha b}\tiny(e , y^\alpha)$
and $F_{\alpha\beta} (y^\alpha)$, we will show that the original models
are equivalent to the $H_4$ and $GL(2 , \mathbb{R})$ WZW models. In this way, the new dual backgrounds for
these WZW models are obtained.

\subsection{\label{subSec.A}The $H_4$ WZW model from T-dualizable $\sigma$-models and its dual pairs}

In this subsection, we obtain two different duals of the $H_4$ WZW model.
In both cases, the original $\sigma$-models (which are  equivalent to the
$H_4$ WZW model) are constructed on  the manifold $M \approx O \times \bf G$ with
${ {\bf G}}=A_2$ acting freely on it, however, the spectator-dependent background matrices are chosen to be
different for each model.\\\\
{\bf Case (1):} In this case, we take the background matrices as
{\small{\begin{eqnarray}\nonumber
E^{+}_{0\;ab} =\left( \begin{array}{cc}
                    0 & e^{y_1}\\
                    e^{y_1} & 0
                      \end{array} \right),~~~
F_{\alpha\beta} =\left( \begin{array}{cc}
                    0 & -1\\
                    -1 & 0
                      \end{array} \right),
\end{eqnarray}}}
\vspace{-2mm}
{\small{\begin{eqnarray}\label{d.4}
F^{+^{(1)}}_{a \beta} =\left( \begin{array}{cc}
                    0 & 0\\
                    e^{y_1} & 0
                      \end{array} \right),~~~
F^{+^{(2)}}_{\alpha b} =\left( \begin{array}{cc}
                    0 & -e^{y_1}\\
                    0 & 0
                      \end{array} \right).
\end{eqnarray}}}
As explained above, the Poisson bracket $\Pi^{ab}(g)$  is zero. By using relations \eqref{a.13}-\eqref{a.16}
one can get the required couplings which where mentioned above. Finally,
using \eqref{d.3} and then \eqref{a.27}, the original $\sigma$-model is found to be of the form
\begin{eqnarray}
S &=& \frac{1}{2} \int d \sigma^+ d \sigma^-~\Big[-\partial_+ y_1 \partial_- y_2-\partial_+ y_2 \partial_- y_1\nonumber\\
&&~~~~~~~~~+ e^{x_1 + y_1} (\partial_+ x_1 \partial_- x_2 +\partial_+ x_2 \partial_- x_1)\nonumber\\
&&~~~~~~~~~+e^{x_1 + y_1}(\partial_+ x_2 \partial_- y_1 - \partial_+ y_1 \partial_- x_2)\Big].\label{d.5}
\end{eqnarray}
By identifying action (\ref{d.5}) with the $\sigma$-model
of the form (\ref{a.3}) one can read off the background matrix.
Thus, the  metric and  antisymmetric tensor field corresponding to the action \eqref{d.5} can be written as
\begin{eqnarray}
d {s}^2 &=&-2 d y_1 d y_2 + 2 e^{x_1 +y_1} d x_1~ d x_2,\label{d.6.1}\\
{B} &=& e^{x_1 + y_1}~d x_2 \wedge d y_1.\label{d.6.2}
\end{eqnarray}
Before proceeding  to construct the dual $\sigma$-model, let  us  discuss the conformal invariance of  the model \eqref{d.5}.
The classical canonical equivalence to the  $\sigma$-models related by PL T-duality was done by Sfetsos in \cite{Sfetsos1} (see, also,
\cite{Sfetsos4}). The canonical transformations are essentially classical and
the quantum equivalence of the two  $\sigma$-models has not yet been revealed. Equivalence
can hold in some special cases but it fails in most cases.
In this respect, checking the equivalence by studying conformal
invariance (the vanishing of the Beta-functions) is important. But, since
after a classical canonical transformation, the equivalence always holds up
to first order in Planck's constant in the semiclassical expansion
(corresponding to one-loop order in $\sigma$-model language), only the two-loop order
is the first real test of quantum equivalence of the
two different $\sigma$-models related by PL T-duality.
For these reasons it is important to check the conformal invariance conditions of our models.

In the $\sigma$-model context, the conformal invariance is
provided by the vanishing of the $B$-functions equations \cite{Tseytlin}, which are equivalent
to the equations of motion of effective action in the string
frame \cite{Metsaev}.
In four dimensions, the low energy string effective action is
\begin{eqnarray}
S_{_{eff}} = \int d^4x \sqrt{-G} e^{-\phi}~ {\cal L}_{_{eff}},\label{d.6.3}
\end{eqnarray}
where $G = \det G_{_{\mu\nu}}$, and ${\cal L}_{_{eff}}$ is given by
\begin{eqnarray}
{\cal L}_{_{eff}} =  {\cal{R}}+ (\nabla \phi)^2- \frac{1}{3} H^2+2\Lambda.\label{d.6.4}
\end{eqnarray}
In this expression,   ${\cal{R}}$ is the scalar curvature of the metric $G_{_{\mu\nu}}$,
and $H_{_{\mu\nu\rho}}$, defined by $H_{_{\mu \nu \rho}}=1/2(\partial_{_\mu} B_{_{\nu \rho}}+
\partial_{_{\nu}} B_{_{\rho\mu}}+\partial_{_{\rho}} B_{_{\mu \nu}})$ is
the torsion (the field strength) of the field  $B_{_{\mu\nu}}$. $\Lambda$ is a
cosmological constant \footnote{In string theory, the cosmological constant term $\Lambda$
is related to the dimension of spacetime, $d$,  and the inverse string
tension by $\Lambda = (d-26)/3\alpha'$, whereas, here in this paper it
is, in some cases,  treated as a free parameter.} which is vanished for critical strings. Our analysis applies
also to non-critical strings, i.e.  when  $\Lambda$ is different from zero.

Consistency of the string theory requires that the action \eqref{a.3}
be defined a conformally invariant quantum field theory.  The conditions for conformal
invariance can be interpreted as field equations for ${G}_{_{\mu \nu}}$,
${B}_{_{\mu \nu}}$ and $\phi$ of the string effective action \cite{{A.Sen},{callan}}.
The vanishing of the one-loop $B$-functions equations gives us the conformal invariance conditions of the
$\sigma$-model \eqref{a.3} up to one-loop order (zeroth
order in the inverse string tension $\alpha'$) \cite{c.hull}. These equations are given by
\begin{subequations}
\begin{eqnarray}
B^{^{G}}_{\mu\nu} :\hspace{-1mm}&&{\cal R}_{{\mu \nu}}-(H^2)_{\mu \nu}+{\nabla}_\mu
{\nabla}_\nu \phi + {\cal O}(\alpha')=0,\label{d.6.1.28}\\
B^{^{B}}_{\mu\nu} :\hspace{-1mm}&&-{\nabla}^\lambda H_{{\lambda \mu \nu}} + H_{{\mu \nu}}^{~\;\lambda}  {\nabla}_\lambda\phi +{\cal O}(\alpha')=0,\label{d.6.1.29}\\
B^{^{\phi}} :\hspace{-1mm}&& \Lambda +\frac{1}{2} {\nabla}^2 \phi -
\frac{1}{2} ({\nabla} \phi)^2+\frac{1}{3} H^2 + {\cal O}(\alpha')=0.~~~~~~~~\label{d.6.1.30}
\end{eqnarray}
\end{subequations}
We have used the conventional notations ${(H^2)}_{_{\mu\nu}}=H_{_{\mu\rho\sigma}}  H^{^{\rho\sigma}}_{_{~~\nu}}$, $H^2=H_{_{\mu\nu\sigma}}
H^{^{\mu\nu\sigma}}$ and $({\nabla} \phi)^2 =\partial_{\mu} \phi ~\partial^{\mu} \phi$.
In the equation \eqref{d.6.1.28}, ${\cal R}_{_{\mu\nu}}$ is the Ricci tensor of the metric
$G_{_{\mu\nu}}$.

The metric \eqref{d.6.1} describes a four-dimensional spacetime of signature $(2 , 2)$.\footnote{
$(2, 2)$-signature often appears in Kleinian geometry as the neutral (- - + +)-signature.
Metrics with $(2, 2)$-signature might seem a purely mathematical problem,
but there are  several physical reasons that motivate this. First of all, two-time physics
(cf. \cite{Bars1} for a review)  has interesting applications in various areas,
like cosmology \cite{Bars2} or M-theory \cite{Bars3}.
Moreover, these metrics are intimately related to
twistor space \cite{Penrose},
which is an important tool in perturbative computations of scattering
amplitudes in gauge theories \cite{Ed. Witten}.}
One quickly  finds that the only non-zero component of ${\cal R}_{_{\mu\nu}}$
is ${\cal R}_{_{y_1 y_1}} =-{1}/{2}$ and then ${\cal R}=0$. Thus, the metric is flat in the sense that its scalar curvature
vanishes \footnote{The metric \eqref{d.6.1} can be considered as the plane-parallel (pp-)wave
in the so-called Rosen coordinates \cite{pp-wave1}. To this end, one can first use the coordinate transformation $e^{x_1} = \theta+\varphi, ~x_2=-\theta+\varphi,~ y_1=u, ~y_2=-v$ to obtain $
d {s}^2 =2 d u ~d v + 2 e^u~(-d\theta^2+ d \varphi^2),$
then, after the change of  the metric signature  by a Wick rotation as $\theta=it$, the resulting metric turns into the pp-wave one
in the  Rosen coordinates.}.  For  the antisymmetric tensor field \eqref{d.6.2}
 one verifies that the only non-zero component of
$H$ is $H_{x_{_1} x_{_2} y_1} =(e^{x_1 + y_1})/2$. It then follows that $H^2 =0$ and the only non-zero component of $(H^2)_{_{\mu\nu}}$ is $(H^2)_{_{y_1y_1}}=-1/2$.
Inserting the above results in the vanishing of the one-loop $B$-functions equations \eqref{d.6.1.28}-\eqref{d.6.1.30},
the conformal invariance conditions up to one-loop order are satisfied with $\Lambda =0$ and the dilaton field
\begin{eqnarray}
\phi &=&\sigma_{_{0}}+ \sigma_{_{1}} y_{_{1}},\label{d.6.1.1}
\end{eqnarray}
where $\sigma_{_{0}}$ and  $\sigma_{_{1}}$ are integration constants.
In addition to the conformal invariance of the model (\ref{d.5}) up to one-loop order, we are interested in
investigating the conformal invariance of the model for higher orders in $\alpha'$.
Instead of this, we show that the model (\ref{d.5})
is equivalent to an exact CFT, namely a WZW model based on a Lie group.
The WZW models represent exact solutions to the string equations of motion to all orders in $\alpha'$.
One can show that under the coordinate transformation
\begin{eqnarray}\label{d.7}
e^{x_1} ~=~ a_+,~~~x_2~=~a_-,~~~y_1~=~ n,~~~y_2~=~ m,
\end{eqnarray}
action \eqref{d.5} turns into
\begin{eqnarray}
S &=& \frac{1}{2} \int d \sigma^+ d \sigma^-~\Big[-\partial_+ n \partial_- m-\partial_+ m \partial_- n\nonumber\\
&&~~~~~~~~~~+ e^{n} (\partial_+ a_+ \partial_- a_- +\partial_+ a_- \partial_- a_+)\nonumber\\
&&~~~~~~~~~~+a_+ e^{n}(\partial_+ a_- \partial_- n - \partial_+ n \partial_- a_-)\Big],\label{d.8}
\end{eqnarray}
which is nothing but the action of WZW model based on the
Lie group $H_4$ \footnote{The WZW model based on the $H_4$
Lie group (a different real form of the $h_4$ Lie algebra of $H_4$) was,
for the first time, introduced by Nappi and Witten \cite{Witten}.} \cite{eghbali11} (cf. Appendix A). Therefore,
the action \eqref{d.5} as an exact CFT describes
string propagation on a four-dimensional manifold with $(2 , 2)$-signature.
We showed that the PL T-duality relates the $H_4$ WZW model to
a $\sigma$-model defined on  $2+2$-dimensional
manifold $M \approx O \times {\bf G}$ only when ${\bf G}$ is the Lie group $A_2$.

To continue, we obtain a new dual background for the $H_4$ WZW model.
This background is obtained from a $\sigma$-model which is constructed on  $2+2$-dimensional manifold
$\tilde {M} \approx O \times \tilde {\bf G}$
with two-dimensional Abelian Lie group ${\tilde {\bf G}}=2A_1$ acting freely on it.
In order to construct the dual model in the form \eqref{a.31} we need to determine the dual couplings.
Making use of \eqref{d.9} and inserting \eqref{d.4} into equations \eqref{a.32}-\eqref{a.35}  they are then read off to be
{\small{\begin{eqnarray}\nonumber
\mathbb{\tilde E}^{ab}=\left( \begin{array}{cc}
                    0 & \frac{1}{e^{y_1}+{\tilde x}_2}\\
                    \frac{1}{e^{y_1}-{\tilde x}_2} & 0
                      \end{array} \right),~~~~~
{\tilde \Phi}_{_{\alpha \beta}}=\left( \begin{array}{cc}
                    0 & -1\\
                    -1 & 0
                      \end{array} \right),
\end{eqnarray}}}
\vspace{-4mm}
{\small{\begin{eqnarray}\label{d.10}
{\tilde \Phi}^{\hspace{-1mm}+^{(1)^{ a}}}_{~~~~\beta}=\left( \begin{array}{cc}
                    \frac{e^{y_1}}{e^{y_1}+{\tilde x}_2} & 0\\
                    0 & 0
                      \end{array} \right),~~~~~~
{{\tilde \Phi}^{+^{{(2)}^{b}}}}_{~\alpha}=\left( \begin{array}{cc}
                    \frac{e^{y_1}}{e^{y_1}-{\tilde x}_2} & 0\\
                    0 & 0
                      \end{array} \right).
\end{eqnarray}}}
Putting these pieces together into \eqref{a.31} and using the fact that the components of the right invariant one-forms on
$2A_1$ are ${\tilde R}_{\pm_{a}}=\partial_{\pm} {{\tilde x}_a}$,
the action of dual $\sigma$-model is obtained to be
{\small \begin{eqnarray}
{\tilde S} &=&\frac{1}{2} \int d \sigma^+ d \sigma^-\Big\{-\partial_+y_1\partial_-y_2 -\partial_+y_2\partial_-y_1~~\nonumber\\
&+&\hspace{-1mm}\frac{1}{{\Delta}}\Big[(e^{y_1}-{\tilde x}_2) \partial_+{{\tilde x}_1}
\partial_-{{\tilde x}_2}+(e^{y_1}+{\tilde x}_2) \partial_+{{\tilde x}_2}
\partial_-{{\tilde x}_1}\nonumber\\
&+&\hspace{-1mm}e^{y_1}(e^{y_1}-{\tilde x}_2) \partial_+{{\tilde x}_1}\partial_-y_1+e^{y_1}(e^{y_1}+{\tilde x}_2)
\partial_+y_1\partial_-{{\tilde x}_1}\Big]\Big\},~~~~~~\label{d.11}
\end{eqnarray}}
where ${\Delta} = e^{2y_1}-{{\tilde x}_2}^2$. Comparing the above action with the $\sigma$-model action of the form \eqref{a.3},
the corresponding line element and antisymmetric field $\tilde B$ take the following forms
\begin{eqnarray}
{d {\tilde s}}^2 &=&-2d y_1~ d y_2+ 2\frac{e^{y_1}}{{\Delta}}\big(d {{\tilde x}_1}~d {{\tilde x}_2}+
e^{y_1}~ d {{\tilde x}_1}~d y_1\big),~~~~\label{d.12}\\
{\tilde B} &=& - \frac{{{\tilde x}_2}}{{\Delta}}\big(d {{\tilde x}_1} \wedge d {{\tilde x}_2}+
e^{y_1}~ d {{\tilde x}_1}  \wedge d y_1\big).\label{d.13}
\end{eqnarray}
The line element \eqref{d.12} is ill defined at the regions ${\tilde x}_2 =  e^{y_1}$ and
${\tilde x}_2 = - e^{y_1}$. We can test whether  there are true singularities
by calculating the scalar curvature, which is, ${\tilde {\cal R}} =0$. Furthermore,
one gets that the only non-zero components of the Ricci tensor and  Riemann tensor field are, respectively,
\begin{eqnarray}
{\tilde {\cal R}}_{_{{\tilde x}_2 y_1}} =\frac{e^{y _1}}{(e^{y _1} - {\tilde x}_2)^2},~
{\tilde {\cal R}}_{_{{y_1 y_1}} }=-\frac{e^{2y _1}+6{\tilde x}_2 e^{y _1}
+ {{\tilde x}_2}^2}{2(e^{y _1} - {\tilde x}_2)^2},~~~\label{d.13.1}
\end{eqnarray}
and
\begin{eqnarray}
{\tilde {\cal R}}_{_{{\tilde x}_1 {\tilde x}_2 {\tilde x}_2 y_1}} &=&\frac{e^{2 y _1}}{(e^{y _1} + {\tilde x}_2)(e^{y _1} - {\tilde x}_2)^3},\nonumber\\
{\tilde {\cal R}}_{_{{\tilde x}_1 y_1 {\tilde x}_2  y_1}} &=&-\frac{e^{y _1} (e^{2y _1} + 6 e^{y _1}{\tilde x}_2 +{{\tilde x}_2}^2)}{4(e^{y _1} + {\tilde x}_2)(e^{y _1} - {\tilde x}_2)^3}.\label{d.13.1.2}
\end{eqnarray}
Then, the other invariant
characteristics of spacetime, such as ${\tilde {\cal R}}_{_{\mu\nu}} {\tilde {\cal R}}^{^{\mu\nu}}$ and
the Kretschmann scalar are found to be  zero. Therefore, the singular points are not the essential singularities,
that is, they can  be removed by an appropriate change of coordinates.
In order to investigate the conformal invariance conditions of the dual model \eqref{d.11}
we look at the vanishing of the one-loop $B$-functions equations \eqref{d.6.1.28}-\eqref{d.6.1.30}.
To this end, we find that the only non-zero component of
$\tilde H$ corresponding to $\tilde B$-field \eqref{d.13}
is ${\tilde H}_{_{{\tilde x}_1 {\tilde x}_2 y_1}}=e^{y _1}/2(e^{y _1} - {\tilde x}_2)^2$;
consequently ${\tilde H}^2 =0$.
Hence,  equations \eqref{d.6.1.28} and \eqref{d.6.1.29} are satisfied by the new dilaton field
\begin{eqnarray}
{\tilde \phi} ~=~ b_{_{0}}+b_{_{1}} y_1- \log \Big(\frac{{\tilde x}_2 - e^{y_1}}{{\tilde x}_2 + e^{y_1}}\Big),\label{d.14}
\end{eqnarray}
where $b_{_{0}}$ and  $b_{_{1}}$ are  integration constants.
The dilatonic contribution, equation \eqref{d.6.1.30}, is  also satisfied if the cosmological constant of the dual theory is left
invariant, that is, ${\tilde \Lambda} = 0$.
Thus, it seems that under the non-Abelian T-duality the cosmological constant
has been restored from the dual model to the original one.

At the end of this subsection let us  discuss the invariance of the string  effective Lagrangians
corresponding to the $\sigma$-models  \eqref{d.5} and \eqref{d.11}. For these models,  the two Lagrangians
${\cal L}_{_{eff}}$ and ${\tilde {\cal L}_{_{eff}}}$ yield the same expression. They are both equal to zero.
The equivalence of Lagrangians holds in spite of the non-vanishing traces
of the structure constants corresponding to the double $({\cal A}_2 , 2{\cal A}_1)$.
Moreover, in the case of this example one can show that the integration weights $\sqrt{-G} e^{-\phi}$
and $\sqrt{- \tilde G} e^{-\tilde \phi}$ are not equal.
The reason behind this can be interpreted in two  ways: firstly, the  dilaton
obtained in \eqref{d.14} does not follow the formula \eqref{a.35.2}.
Secondly, due to the particularity of our model, there is a possibility of absorbing the anomalous terms into
dilaton shift which is the same as a diffeomorphism transformation.
An analysis similar to this has been carried out in Ref. \cite{Balog} in a strict field theory sense, regardless of the
relationship between $\sigma$-models and string theory effective actions.
Notice that the equations \eqref{a.35.1} and \eqref{a.35.2} are  the only transformations which lead to a proportionality
between the integration weights $\sqrt{-G} e^{-\phi}$ and $\sqrt{- \tilde G} e^{-\tilde \phi}$ \cite{N.Mohammedi}.\\\\
{\bf Case (2):}{~\it The self-duality of the $H_4$ WZW model}\\

The self-duality of the WZW model under PL T-duality, as well as the $SU(N)$ WZW model, has already been discussed in  \cite{Klim3}.
It turns out that the dual to the WZW model is again the same WZW model.
Here we shall show that the  $H_4$ WZW model is self-dual.
Let us now choose the spectator-dependent background  matrices as
{\small{\begin{eqnarray}
E^{+}_{0\;ab}&=&\left( \begin{array}{cc}
                    0 & -{y_1}\\
                    {y_1} & 0
                      \end{array} \right),~~~
F^{+^{(1)}}_{a \beta}=\left( \begin{array}{cc}
                    0 & -1\\
                    1 & 0
                      \end{array} \right),\nonumber\\
F^{+^{(2)}}_{\alpha b}&=&\left( \begin{array}{cc}
                    0 & 1\\
                    -1 & 0
                      \end{array} \right),~~~~~~~F_{\alpha\beta}=0.\label{d.14.1}
\end{eqnarray}}}
Then, using the fact that $\Pi(g)=0$ and utilizing formulas \eqref{a.13}-\eqref{a.16} together with \eqref{d.3} and \eqref{a.27}, the original $\sigma$-model is, in this case, obtained to be of the form
\begin{eqnarray}
S &=& \frac{1}{2} \int d \sigma^+ d \sigma^-~\Big[- y_{_{1}} e^{x_{_{1}}}(\partial_+ x_{_{1}}
\partial_- x_{_{2}} - \partial_+ x_{_{2}} \partial_- x_{_{1}})\nonumber\\
&&~~~~~~~~~~~~~~~~~+ e^{{x_{_1}}} (\partial_+ y_{_{1}} \partial_- x_{_{2}} +\partial_+ x_{_{2}} \partial_- y_{_{1}})\nonumber\\
&&~~~~~~~~~~~~~~~~~-\partial_+ x_{_{1}} \partial_- y_{_{2}}-\partial_+ y_{_{2}} \partial_- x_{_{1}}
\Big].\label{d.14.2}
\end{eqnarray}
As it is seen, the action  \eqref{d.14.2} is indeed identical to the action of  $H_4$ WZW model (cf. Appendix A).

Analogously to Case (1), the dual model is constructed on the
manifold ${\tilde M } \approx O \times \tilde {\bf G}$. The Poisson bracket on
$\tilde {\bf G}$ is given by  formula \eqref{d.9}, and thus by inserting \eqref{d.14.1} into equations
\eqref{a.32}-\eqref{a.35}  the dual couplings are computed to be
{{\begin{eqnarray}\nonumber
\mathbb{\tilde E}^{ab}=\frac{1}{y_{_1}+{\tilde x}_{_2}}\left( \begin{array}{cc}
                    0 & 1\\
                    -1 & 0
                      \end{array} \right),~
{\tilde \Phi}_{_{\alpha \beta}}=\frac{1}{y_{_1}+{\tilde x}_{_2}}\left( \begin{array}{cc}
                    0 & -1\\
                    1 & 0
                      \end{array} \right),
\end{eqnarray}}}
\vspace{-2mm}
{\small{\begin{eqnarray}\label{d.14.3}
{\tilde \Phi}^{\hspace{-1mm}+^{(1)^{ a}}}_{~~~~\beta}=\frac{1}{y_{_1}+{\tilde x}_{_2}}\left( \begin{array}{cc}
                    1 & 0\\
                    0 & 1
                      \end{array} \right),~
{{\tilde \Phi}^{+^{{(2)}^{b}}}}_{~\alpha}=\frac{1}{y_{_1}+{\tilde x}_{_2}}\left( \begin{array}{cc}
                    1 & 0\\
                    0 & 1
                      \end{array} \right).~~~~
\end{eqnarray}}}
Finally, inserting these into  formula \eqref{a.31} the dual $\sigma$-model is obtained to be in the following form
\begin{eqnarray}
{\tilde S} &=& \frac{1}{2} \int d \sigma^+ d \sigma^-
\frac{1}{(y_{_1}+{\tilde x}_{_2})}\Big[\partial_+{\tilde x}_{_{1}}~\partial_-y_{_{1}} +\partial_+y_{_{1}}~\partial_-{\tilde x}_{_{1}}\nonumber\\
&&~~~~+\partial_+{\tilde x}_{_{2}}~\partial_-y_{_{2}} +\partial_+y_{_{2}}~\partial_-{\tilde x}_{_{2}}
+\partial_+{\tilde x}_{_{1}}~\partial_-{\tilde x}_{_{2}}\nonumber\\
&&~~~~-\partial_+{\tilde x}_{_{2}}~\partial_-{\tilde x}_{_{1}}-
\partial_+y_{_{1}}~\partial_-y_{_{2}} + \partial_+y_{_{2}}~\partial_-y_{_{1}}\Big].\label{d.14.4}
\end{eqnarray}
The line element and antisymmetric tensor field corresponding to action  \eqref{d.14.4} can be cast in the forms
\begin{eqnarray}
{d {\tilde s}}^2 &=& \frac{2}{(y_{_1}+{\tilde x}_{_2})} \big(d {\tilde x}_{_{1}}~
d {y}_{_{1}}+ d {\tilde x}_{_{2}}~ d {y}_{_{2}}\big),\label{d.14.5}\\
{\tilde B} &=& \frac{1}{(y_{_1}+{\tilde x}_{_2})} \big(d {\tilde x}_{_{1}} \wedge
d {\tilde x}_{_{2}}-  d {y}_{_{1}}  \wedge d {y}_{_{2}}\big).\label{d.14.6}
\end{eqnarray}
By using \eqref{d.14.5} and \eqref{d.14.6}, the conformal invariance conditions of the model
\eqref{d.14.4} are satisfied with zero cosmological constant and  dilaton field
that supports the dual background is found to be
\begin{eqnarray}
{\tilde \phi} ~=~c_{_{0}} + {c_{_{1}}} \log (y_{_1}+{\tilde x}_{_2}),\label{d.14.7}
\end{eqnarray}
where $c_{_{0}}$ and $c_{_{1}}$ are the  constants of integration. On the one hand, it is interesting to note that
if we write \eqref{d.14.7} as $e^{-{\tilde \phi}} =\varrho_{_0} /(y_{_1}+{\tilde x}_{_2})$, then metric \eqref{d.14.5}
may be expressed as \footnote{Here we have set $\varrho_{_0}=1$.}
\begin{eqnarray}
{\tilde G}_{_{\mu\nu}}=e^{-{\tilde \phi}} ~{\hat \eta}_{_{\mu\nu}},\label{d.14.8}
\end{eqnarray}
in which  ${\tilde \eta}_{_{\mu\nu}}= 2 d {\tilde x}_{_{1}}~
d {y}_{_{1}}+ 2d {\tilde x}_{_{2}}~ d {y}_{_{2}} $.   The formula \eqref{d.14.8}
indicates a conformal transformation \footnote{The conformal transformations shrink or stretch the distances
between the two points described by the same coordinate system $x^\mu$ on the manifold $M$,
but they preserve the angles between vectors  which lead to a conservation of the (global) causal structure of the manifold \cite{Hawking}.}
between the dual metric ${\tilde G}_{_{\mu\nu}}$ and flat metric ${\tilde \eta}_{_{\mu\nu}}$, and
$e^{-{\tilde \phi}}$ is  a smooth, non-vanishing function of the spacetime  which is called a conformal factor.
We note that the conformal transformations do change geometry and they are entirely different from
coordinate transformations. This is crucial since conformal transformations may lead to a different physics \cite{Hawking}.
If we use the coordinate transformation
\begin{eqnarray}\label{d.14.9}
{\tilde x}_{_1} =m,~{\tilde x}_{_2}=a_{_+},~y_1= e^{-n}-a_+,~y_2= a_-+m,
\end{eqnarray}
then the dual background can be cast to \cite{eghbali11}
\begin{eqnarray}
{d {\tilde s}}^2 &=& -2d n ~ d m+ 2 e^{n} d a_{_+} d a_{_-},\nonumber\\
{\tilde B} &=& -a_{_+} e^{n} ~d n  \wedge d a_{_-}.\label{d.14.10}\\
{\tilde \phi}& =& \varsigma_{_{0}}+ \varsigma_{_{1}} n,\nonumber
\end{eqnarray}
where $\varsigma_{_{0}}$ and $\varsigma_{_{1}}$ are arbitrary constants.
Here we have ignored the total derivative terms that appeared in the
${\tilde B}$-field part.
Indeed, the solution \eqref{d.14.10} is identical to the background of the original  $\sigma$-model action \eqref{d.14.2}.
{\it Thus, we showed that the $H_4$ WZW model does remain invariant under the non-Abelian T-duality transformation, that is, the model is self-dual.}

\subsection{\label{subSec.B}The $GL(2,\mathbb{R})$ WZW model from T-dualizable $\sigma$-models and its dual pair}

We shall show that the original $\sigma$-model
\eqref{a.27} on the $2+2$-dimensional manifold $M \approx O \times \bf G$ can be  equalled to the $GL(2,\mathbb{R})$ WZW model.
Similar to previous examples, the isometry group $\bf G$ that is being dualized is $A_2$.
The only difference is in choosing the spectator-dependent background  matrices.
In this regard, the non-Abelian T-dual geometry of the $GL(2,\mathbb{R})$ WZW model is determined.

\subsubsection{The original $\sigma$-model as the $GL(2,\mathbb{R})$ WZW model}

 Here, we choose
{\small{\begin{eqnarray}\nonumber
E^{+}_{0\;ab} = \left( \begin{array}{cc}
                    0 & \frac{1}{2}e^{-2y_1}\\
                   \frac{1}{2}e^{-2y_1} & 0
                      \end{array} \right),~~~~
F_{\alpha\beta}  = \left( \begin{array}{cc}
                    1 & 0\\
                    0 & b
                      \end{array} \right),
\end{eqnarray}}}
\vspace{-2mm}
{\small{\begin{eqnarray}\label{d.15}
F^{+^{(1)}}_{a \beta} = \left( \begin{array}{cc}
                    0 & 0\\
                    -e^{-2y_1} & 0
                      \end{array} \right),~~~~
F^{+^{(2)}}_{\alpha b} =\left( \begin{array}{cc}
                    0 & e^{-2y_1}\\
                    0 & 0
                      \end{array} \right),
\end{eqnarray}}}
where $b$ is a non-zero real constant. Inserting relations \eqref{d.15} into \eqref{a.13}-\eqref{a.16} and then  using
\eqref{d.3} together with \eqref{a.27}, the original $\sigma$-model is worked out to be
\begin{eqnarray}
S &=& \frac{1}{2} \int d \sigma^+ d \sigma^-~\Big[\partial_+ y_1 \partial_- y_1+b ~\partial_+ y_2 \partial_- y_2\nonumber\\
&&~~~~~~~~~+ \frac{1}{2} e^{x_1 -2y_1} (\partial_+ x_1 \partial_- x_2
+\partial_+ x_2 \partial_- x_1)\nonumber\\
&&~~~~~~~~~+e^{x_1 -2 y_1}(\partial_+ y_1  \partial_- x_2 - \partial_+ x_2 \partial_- y_1)\Big].~~\label{d.16}
\end{eqnarray}
Now, if one uses the following coordinate transformation
\begin{eqnarray}\label{d.17}
e^{x_1} ~=~ \theta_{_{-}},~~~x_2~=~\theta_{_{+}},~~~y_1~=~ \theta_{_{3}},~~~y_2~=~ \theta,
\end{eqnarray}
then, action \eqref{d.16} becomes
\begin{eqnarray}
S &=& \frac{1}{2} \int d \sigma^+ d \sigma^-~\Big[b ~ \partial_{_{+}} \theta ~\partial_{_{-}} \theta +
\partial_{_{+}} \theta_{_{3}} \partial_{_{-}} \theta_{_{3}} \nonumber\\
&&~~~~~~~~~+ \frac{1}{2}e^{-2 \theta_{_{3}}} (\partial_{_{+}}
\theta_{_{-}} \partial_{_{-}} \theta_{_{+}} + \partial_{_{+}}
\theta_{_{+}} \partial_{_{-}} \theta_{_{-}}) \nonumber \\
&&~~~~~~~~~+
\theta_{_{-}} e^{-2 \theta_{_{3}}} (\partial_{_{+}} \theta_{_{3}} \partial_{_{-}} \theta_{_{+}} -
\partial_{_{+}} \theta_{_{+}} \partial_{_{-}} \theta_{_{3}})\Big].~~\label{d.18}
\end{eqnarray}
Using the integration by parts over the fourth term of action, it is concluded that the action (\ref{d.18})
is nothing but the action of WZW model based on the Lie group $GL(2,\mathbb{R})$ (cf.  Appendix A).
Hence, the original $\sigma$-model (\ref{d.16}) can be described as an exact CFT.

The line element and the antisymmetric tensor field corresponding to  action (\ref{d.16}) are, respectively, given by
\begin{eqnarray}
d {s}^2 &=& {d y_1}^2  + b ~{d y_2}^2 + e^{x_1 -2 y_1} d x_1~ d x_2,\label{d.19}\\
{B} &=& - e^{x_1 -2 y_1}~d x_2 \wedge d y_1.\label{d.20}
\end{eqnarray}
To better understand action \eqref{d.16} we diagonalize the corresponding metric. Let
\begin{eqnarray}
e^{x_1} &=& \frac{1}{l}({t}- l {\varphi}),~
x_2  = -l({t}+ l {\varphi}),\nonumber\\
e^{y_1}  &=&\frac{l}{{r}},~~~~~~~~~~~y_{_{2}} ~=~z, ~\label{d.21}
\end{eqnarray}
where $l$ is constant with the dimension of length. Then, the metric \eqref{d.19} and the field strength corresponding to the $B$-field
\eqref{d.20} shall, respectively, become
\begin{eqnarray}
d {s}^2 &=& -\frac{r^2}{l^2} {d t}^2  + {r^2} {d \varphi}^2 + \frac{1}{r^2} {d r}^2 +b~ {d z}^2,\label{d.22}\\
{H} &=& -\frac{ r}{l}  dt \wedge d \varphi  \wedge  d r.\label{d.23}
\end{eqnarray}
The metric \eqref{d.22}  describes a four-dimensional
Lorentz-signature spacetime if $b$ is considered positive and when $b$ is negative, the metric has $(2 , 2)$-signature.
One immediately finds that the scalar curvature  of the metric is ${\cal R}=-{6}$,
and the non-zero components of ${\cal R}_{_{\mu \nu}}$ are
${\cal R}_{_{nn}}=-2 {g}_{_{nn}}$ where  $n=(t, \varphi, r)$.
Since ${\cal R}_{_{zz}}=0$, this spacetime cannot be described as an $AdS_{4}$ space.
On the other hand, the metric is a direct product of $\mathbb{R}$ associated with the
coordinate $z$ and the three-dimensional metric of $(t, \varphi, r)$, which is nothing but the $AdS_{3}$ space. Hence, the metric
\eqref{d.22} corresponds to $AdS_{3} \times \mathbb{R}$. Furthermore,  using \eqref{d.22} and \eqref{d.23}
it is concluded that the only non-zero components of $(H^2)_{_{\mu\nu}}$ are  $(H^2)_{_{tt}}=(2r^2)/l^2, (H^2)_{_{\varphi\varphi}}=-2r^2$ and
$(H^2)_{_{rr}}=-2/r^2$.
Finally, one verifies the equations
\eqref{d.6.1.28} and \eqref{d.6.1.29} with the following  dilaton field
\begin{eqnarray}
\phi &=&\zeta_{_{0}}+ \zeta_{_{1}} z,\label{d.24}
\end{eqnarray}
where $\zeta_{_{0}}$ and  $\zeta_{_{1}}$ are integration  constants.
Also, computing $H^2 = -6$ the dilatonic contribution in equation \eqref{d.6.1.30} is  satisfied
provided that ${\zeta_{_{1}}}^2  = b(2\Lambda -4)$.
Notice that the metric \eqref{d.22} has no horizon and no curvature singularity. Indeed,
this solution is everywhere regular including $r=0$.
Consider now two killing vectors
$\partial/{\partial t}$ and $\partial/{\partial {\varphi}}$ corresponding to the time translational
and the rotational isometries of the metric \eqref{d.22}, respectively. The killing field
$\partial/{\partial t}$ becomes null at $r=0$ and it is time-like for the whole range $r>0$, while
the killing field   $\partial/{\partial {\varphi}}$ is everywhere space-like except for  $r=0$. In addition,
there is another  killing field  such as $(1/b)~\partial/{\partial z}$ so that it is time-like for $b<0$, and remains
space-like for $b>0$.

\subsubsection{The dual $\sigma$-model}

Similar to the construction of dual $\sigma$-models for $H_4$ WZW, the dual manifold is, here, assumed to be
$\tilde {M} \approx O \times \tilde {\bf G}$ in which ${\tilde {\bf G}}=2A_1$. So, the Poisson structure on
$2A_1$ does follow the relation \eqref{d.9}. In order to obtain the dual $\sigma$-model
for $GL(2,\mathbb{R})$ WZW, we use the action
\eqref{a.31}. The dual coupling matrices can be obtained by
inserting \eqref{d.9} and \eqref{d.15} into \eqref{a.32}-\eqref{a.35}.
They are then read
{\small{\begin{eqnarray}\nonumber
\mathbb{\tilde E}^{ab}=\left( \begin{array}{cc}
                    0 & \frac{1}{{\tilde x}_2 +\frac{1}{2}e^{-2y_1}}\\
                    \frac{1}{-{\tilde x}_2+\frac{1}{2}e^{-2y_1}} & 0
                      \end{array} \right),~
{\tilde \Phi}_{_{\alpha \beta}} =\left( \begin{array}{cc}
                    1 & 0\\
                    0 & b
                      \end{array} \right),
\end{eqnarray}}}
\vspace{-2mm}
{\small{\begin{eqnarray}\label{d.25}
{\tilde \Phi}^{\hspace{-1mm}+^{(1)^{ a}}}_{~~~~\beta} =\left( \begin{array}{cc}
                    -\frac{e^{-2y_1}}{{\tilde x}_2+\frac{1}{2}e^{-2y_1}} & 0\\
                    0 & 0
                      \end{array} \right),~
{{\tilde \Phi}^{+^{{(2)}^{b}}}}_{~\alpha} =\left( \begin{array}{cc}
                    \frac{e^{-2y_1}}{{\tilde x}_2-\frac{1}{2}e^{-2y_1}} & 0\\
                    0 & 0
                      \end{array} \right).~~
\end{eqnarray}}}
Finally, the dual $\sigma$-model to the $GL(2,\mathbb{R})$ WZW model is found  to be
\begin{widetext}
{\begin{eqnarray}
{\tilde S} &=& \frac{1}{2} \int d \sigma^+ d \sigma^-
\Big\{\partial_+y_1~\partial_-y_1 +b~\partial_+y_2~\partial_-y_2+\frac{1}{{\bar{\Delta}}}\Big[(\frac{1}{2}e^{-2y_1}-{\tilde x}_2) \partial_+{{\tilde x}_1}~
\partial_-{{\tilde x}_2}+(\frac{1}{2}e^{-2y_1}+{\tilde x}_2) \partial_+{{\tilde x}_2}~
\partial_-{{\tilde x}_1}\nonumber\\
&&~~~~~~~~~~~~~~~~~~~~~-e^{-2y_1}(\frac{1}{2}e^{-2y_1}-{\tilde x}_2) \partial_+{{\tilde x}_1}~\partial_-y_1
-e^{-2y_1}(\frac{1}{2}e^{-2y_1}+{\tilde x}_2)
\partial_+y_1~\partial_-{{\tilde x}_1}\Big]\Big\},\label{d.26}
\end{eqnarray}}
\end{widetext}
where ${\bar{\Delta}} = \frac{1}{4}e^{-4y_1}-{{\tilde x}_2}^2$. The line element and antisymmetric field corresponding
to this action may be expressed as
\begin{eqnarray}
{d {\tilde s}}^2 &=&{d y_{_{1}}}^2 + b~ {d y_{_{2}}}^2\nonumber\\
~~~~~&&+ \frac{e^{-2y_{_{1}}}}{{\bar{\Delta}}}\big(d {{\tilde x}_1}~d {{\tilde x}_2}
-e^{-2y_{_{1}}}~ d {{\tilde x}_1}~d y_{_{1}}\big),\label{d.27}\\
{\tilde B} &=& - \frac{{{\tilde x}_2}}{{\bar{\Delta}}}\big(d {{\tilde x}_1} \wedge d {{\tilde x}_2}-
e^{-2y_{_{1}}}~ d {{\tilde x}_1}  \wedge d y_{_{1}}\big).\label{d.28}
\end{eqnarray}
The scalar curvature of the metric is
\begin{eqnarray}
{\tilde {\cal R}} ~=~ - \frac{2(11 e^{-4y_{_{1}}} +28 {\tilde x}_2 e^{-2y_{_{1}}} +12 {{\tilde x}_2}^2)}{(e^{-2y_{_{1}}} - 2{\tilde x}_2)^2}.\label{d.29}
\end{eqnarray}
As it can be seen from formulas \eqref{d.27} and \eqref{d.29}, the
region ${\tilde x}_2 = \frac{1}{2} e^{-2y_{_{1}}}$ is a true curvature singularity
(in what follows we will discuss the structure and  asymptotic nature of the dual
spacetime  including the horizon and singularity).
We deduce that the only non-zero component of the field strength corresponding to the $\tilde B$-field \eqref{d.28} is
${\tilde H}_{_{{\tilde x}_1 {\tilde x}_2 y_{_{1}}}}=- (2e^{-2y_{_{1}}})/(e^{-2y_{_{1}}} - 2{\tilde x}_2)^2$; consequently
${\tilde H}^2=-6(e^{-2y_{_{1}}} + 2{\tilde x}_2)^2/(e^{-2y_{_{1}}} - 2{\tilde x}_2)^2$.
Thus, equations  \eqref{d.6.1.28} and \eqref{d.6.1.29} are satisfied by the new dilaton field
\begin{eqnarray}
{\tilde \phi} ~=~ \lambda_{_{0}} + \lambda_{_{1}} y_{_{2}} + \log \Big(\frac{2{\tilde x}_2 + e^{-2y_{_{1}}}}{2{\tilde x}_2 - e^{-2y_{_{1}}}}\Big),\label{d.30}
\end{eqnarray}
where $\lambda_{_{0}}$  and $\lambda_{_{1}}$ are arbitrary constants.  Also, the dilatonic contribution  in  \eqref{d.6.1.30}
is vanished if the cosmological constant of the dual theory does satisfy in ${\lambda_{_{1}}}^2  = b(2{\tilde \Lambda} -4)$.

As in the first example of subsection A, we now discuss the presence of
anomalous terms breaking the proportionality between the original and dual string effective actions.
The string effective Lagrangians corresponding to the $\sigma$-models \eqref{d.16} and \eqref{d.26} are
found to be
\begin{eqnarray}
{\cal L}_{_{eff}} &=& 4 \Lambda -8, \label{d.30.1}\\
{\tilde {\cal L}_{_{eff}}} &=&4 {\tilde \Lambda} -8 -\frac{64 {\tilde x}_2 e^{2y_{_{1}}}}{(1-2{\tilde x}_2 e^{2y_{_{1}}})^2}.  \label{d.30.2}
\end{eqnarray}
The last term of equation \eqref{d.30.2} is not invariant under PL T-duality transformation and therefore
the two Lagrangians  ${\cal L}_{_{eff}}$ and ${\tilde {\cal L}_{_{eff}}}$ are not equal. This anomaly is due to the non-vanishing
traces of the structure constants of the double $({\cal A}_2 , 2{\cal A}_1)$; furthermore,
the dilaton field obtained in \eqref{d.30} does not follow the transformation \eqref{a.35.2}.

The dilaton field \eqref{d.30} is well behaved
for the ranges ${\tilde x}_2 < -\frac{1}{2} e^{-2y_{_{1}}} $ and ${\tilde x}_2 > \frac{1}{2} e^{-2y_{_{1}}}$.
We also note that a  dilaton field can easily be found for the range $-\frac{1}{2} e^{-2y_{_{1}}} <  {\tilde x}_2 < \frac{1}{2} e^{-2y_{_{1}}}$
by shifting $\lambda_{_{0}}$ by an imaginary constant ($\lambda_{_{0}}\rightarrow \lambda_{_{0}} +i\pi$).

For the range ${\tilde x}_2 <-\frac{1}{2} e^{-2y_{_{1}}}$ we consider ${\tilde x}_2 +\frac{1}{2} e^{-2y_{_{1}}} = - e^X$. Then, we introduce
the following coordinate transformation
\begin{eqnarray}
{{\tilde x}_1} &=&  Y +\frac{1}{2} (W+e^{^{W}}),~~{{\tilde x}_2} = -e^X (1+\frac{e^{^{-W}}}{2}),\nonumber\\
{{y}_{_{1}}} &=& \frac{1}{2} (W-X),~~~~ ~~~~~~y_2 = V.  \label{d.31}
\end{eqnarray}
Under this transformation, the dual background now looks as follows
\begin{eqnarray}
{d {\tilde s}}^2 &=&b ~d {V}^2 + \frac{1}{4} (d W^2+ d X^2) + \frac{1}{e^W+1} ~d X d Y,~~~\label{d.32}\\
{\tilde B} &=&  -\frac{2 e^{^{W}}+1}{2(e^{^{W}}+1)} ~ d X \wedge d Y,\label{d.33}\\
{\tilde \phi} &=& \lambda_{_{0}}+ \lambda_{_{1}} V + \log \big(\frac{e^{^{W}}}{e^{^{W}}+1}\big).~~~~\label{d.34}
\end{eqnarray}
Here we have ignored the terms concerning $\tilde B$-field which are
contributed to the Lagrangian as the total derivatives.
Notice that there is no singularity for the metric \eqref{d.32}.  In fact, this was expected since
the solutions \eqref{d.27}, \eqref{d.28} and \eqref{d.30} are, in this case, defined only for the range ${\tilde x}_2 +\frac{1}{2} e^{-2y_{_{1}}} <0$.
As explained above, the true singularity of the metric  \eqref{d.27} occurs  at ${\tilde x}_2 = \frac{1}{2} e^{-2y_{_{1}}}$,
a region which  is located out of the range ${\tilde x}_2 +\frac{1}{2} e^{-2y} <0$.
The background \eqref{d.32}-\eqref{d.34} can be simplified by performing a coordinate transformation.
Let us now consider the transformation $e^{^{W}} = {1}/{(r -1)}$ so that it requires that $1 <{r}< \infty$. In addition,
we introduce the following linear transformation
\begin{eqnarray}
X =  -2 (t +\frac{x}{\sqrt{3}}), ~~~~ Y = (t - \frac{x}{\sqrt{3}}),~~~~V = z.\label{x.5}
\end{eqnarray}
By applying the above transformation to the solutions \eqref{d.32}, \eqref{d.33} and \eqref{d.34}, one obtains
the forms of the dual spacetime metric, antisymmetric field strength and dilaton field in new coordinate base $\{t, x, r, z\}$ as
\begin{eqnarray}
{d {\tilde s}}^2 &=& - (1-\frac{2}{r}) d t^2 +  (1-\frac{2}{3 r}) d x^2 \nonumber\\
&&~~+\frac{2}{\sqrt{3}}~ d t d x +(1-\frac{1}{r})^{-2} ~ \frac{ d r^2}{4 r^2}+b ~dz^2,~\label{x.6}\\
{{\tilde H}_{r t x}} &=& \frac{1}{\sqrt{3}~ r^2},\label{x.7}\\
{\tilde \phi} &=& \lambda_{_{0}}+\lambda_{_{1}}~z - \log r. \label{x.8}
\end{eqnarray}
We note that this solution is valid only for the range ${\tilde x}_2 +\frac{1}{2} e^{-2y_{_{1}}} <0$ or $1 <{r}< \infty$. In
order to have a solution with the range $0 <{r}< 1$ we have to look at the
second case, where ${\tilde x}_2 - \frac{1}{2} e^{-2y_{_{1}}}>0$. In this case it is  assumed that ${\tilde x}_2 - \frac{1}{2} e^{-2y_{_{1}}} = e^{X}-e^{-2y_{_{1}}}$ for which
$X+2y_{_{1}} >0 $. Analogously, we introduce the  transformation
\begin{eqnarray}
{{\tilde x}_1} &=&  Y -\frac{1}{2} (e^{^{W}}-W),~~~{{\tilde x}_2} = e^X (1-\frac{e^{^{-W}}}{2}),\nonumber\\
y_{_{1}} &=& \frac{1}{2} (W-X),~~~~~~~~~~ y_{_{2}} = V,\label{x.9}
\end{eqnarray}
in which $W =X+2y_{_{1}}>0$, i.e., $e^W >1$. We then define the transformation $e^{^{W}} = {1}/{(1-{r})}$ so that it requires that $0 <{r}< 1$.
Using these results and also utilizing the linear transformation \eqref{x.5} one concludes that the solution given by
equations \eqref{d.27}, \eqref{d.28} and  \eqref{d.30} is nothing but the solution given by \eqref{x.6}-\eqref{x.8}.
Thus, the obtained  solutions to both the valid ranges ${\tilde x}_2 +\frac{1}{2} e^{-2y_{_{1}}} <0$
and ${\tilde x}_2 - \frac{1}{2} e^{-2y_{_{1}}}>0$ can be expressed as a solution in the form of equations \eqref{x.6}-\eqref{x.8} only with
$0 <{r}< \infty$. Analogously, for the range $- \frac{1}{2} e^{-2y_{_{1}}} <{\tilde x}_2 < \frac{1}{2} e^{-2y_{_{1}}}$ one can
consider ${\tilde x}_2 +\frac{1}{2} e^{-2y_{_{1}}} =  e^X$ to obtain the same results presented in \eqref{x.6}-\eqref{x.8}.

One can simply check that the solution  \eqref{x.6}-\eqref{x.8} does satisfy the equations  \eqref{d.6.1.28}-\eqref{d.6.1.30}.
Considering this solution for the whole spacetime, $0 <{r}< \infty$,  one sees that
the metric components \eqref{x.6} are ill behaved at $r =0$  and $r = 1$.
Looking at the scalar curvature, which is
${\cal {\tilde R}} ~=~ {2(4 r -7)}/{{r}^2}$,
we find that $r =0$ is a curvature singularity. Notice that the
singularity at  $r =0$ corresponds to the same true
singularity at the region ${\tilde x}_2 = \frac{1}{2} e^{-2y_{_{1}}}$
which mentioned above.
We furthermore  see that $r = 1$ is also an event horizon.
The cross term  appeared in the metric is constant and thus for large $r$ one can show that the metric is asymptotically
flat. For large $r$ the metric \eqref{x.6} approaches the following asymptotic solution
\begin{eqnarray}
{d {\tilde s}}^2 ~=~ -  d t^2 +   d x^2 +
\frac{2}{\sqrt{3}}~ d t d x + \frac{ d r^2}{4 r^2}+b ~dz^2.\label{x.10}
\end{eqnarray}
Performing a convenient coordinate transformation, the metric \eqref{x.10}  can be simply diagonalized.
We also note that the sign of $b$ changes the signature of metric.
If we introduce the new coordinates $(\hat{t}, \hat{x}, \hat{r}, \hat{z})$ by the transformation
\begin{eqnarray}
r =  e^{2\hat{r}},~~~~~t = \frac{\sqrt{3}}{2} \hat{t},~~~~~~
x = \hat{x} - \frac{\hat{t}}{2} ,~~~~ z =  \frac{\hat{z}}{\eta},\label{x.11}
\end{eqnarray}
then, \eqref{x.10} will become
\begin{eqnarray}
{d {\tilde s}}^2 ~=~ \left\{\begin{array}{ll} - d \hat{t}^2 +   d \hat{x}^2 +{ d\hat{ r}^2}+d\hat{z}^2 &{\rm for} \;b={\eta}^2\label{x.12}\\
- d \hat{t}^2 +   d \hat{x}^2 +{ d\hat{ r}^2} - d\hat{z}^2 &{\rm for} \;b=-{\eta}^2\end{array} \right..
\end{eqnarray}
As it is seen for $b>0$ the  metric has $(1 , 3)$-signature, while the signature is $(2 , 2)$ when  $b$ is negative.
{\it Thus, we have  shown that the non-Abelian T-duality
transformation (here as the PL T-duality on a semi-Abelian double)
changes the asymptotic behavior of solutions from $AdS_{3} \times \mathbb{R}$ to flat space.}

\subsection{\label{subSec.C}The non-Abelian T-duality of  non-critical Bianchi type III string
cosmological model (the $GL(2,\mathbb{R})$ WZW model)}

In this subsection, we show that the non-critical Bianchi type III string
cosmology solution with a non-vanishing field strength
and an appropriate dilaton field can be described by the $GL(2,\mathbb{R})$ WZW model.
In fact, we shall obtain the $GL(2,\mathbb{R})$ WZW model from a  T-dualizable $\sigma$-model
constructed on a $3+1$-dimensional manifold $M \approx O \times \bf G$, in which $\bf G$  is
three-dimensional decomposable Lie group $A_2 \oplus A_1$ acting freely on $M$.
In this case, the non-Abelian T-duality of the model is studied here.
The dual Lie group $\tilde {\bf G}$ is considered to be three-dimensional Abelian Lie group $3 A_1$.
We note that the Lie algebra  ${\cal A}_2 \oplus {\cal A}_1$ is isomorphic to the Lie algebra of Bianchi type III.
Hence, six-dimensional Lie algebra of the Drinfeld double $({\cal A}_2 \oplus {\cal A}_1 , 3{\cal A}_1)$ is defined by the following commutation relations:
\begin{eqnarray}\label{x.13}
[T_1 , T_2] &=& T_2,~~~~[T_3~ , ~.]~=~0,~~[T_1 ~, ~{\tilde T}^2]=-{\tilde T}^2,\nonumber\\
{[T_2 ~, ~{\tilde T}^2]} &=& {\tilde T}^1,~~~~[{\tilde T}^3 , ~.]~=~0.
\end{eqnarray}
Taking a convenient element of the Lie group $A_2 \oplus A_1$ such as $g=e^{(\ln x_1) T_1}~e^{x_2 T_2}~e^{x_3 T_3}$
we immediately find that $R_{\pm}^1= {\partial_{\pm} x_{_{1}}}/{x_{_{1}}}$,
$R_{\pm}^2= {x_{_{1}}} \partial_{\pm} x_{_{2}}$, $R_{\pm}^3= \partial_{\pm} x_{_{3}}$. In order to study the
non-Abelian T-duality of  non-critical Bianchi type III string
cosmological model, we consider the orbit $O$ as a one-dimensional space with time coordinate  $y^{\alpha}=\{t\}$.
Now, one can choose the spectator-dependent background  matrices as
{\small{\begin{eqnarray}\nonumber
E^{+}_{0\;ab} =\left( \begin{array}{ccc}
                    0 & -\frac{a_{_{0}}^2}{2}e^{-2t} & 0\\
                   -\frac{a_{_{0}}^2}{2}e^{-2t} & 0 &0\\
                   0 & 0 &b
                      \end{array} \right),~~F^{+^{(1)}}_{a \beta}=\left( \begin{array}{c}
                    0 \\
                    a_{_{0}}^2~e^{-2t} \\
                    0
                      \end{array} \right),
\end{eqnarray}}}
\vspace{-3mm}
{\small{\begin{eqnarray}\label{x.14}
F^{+^{(2)}}_{\alpha b}=\left( \begin{array}{ccc}
                    0 & -a_{_{0}}^2~e^{-2t} &0
                      \end{array} \right),~ ~~~~~~~~~~~~~~   F_{\alpha\beta} =-a_{_{0}}^2,~~~~~~~~
\end{eqnarray}}}
for some constants $a_{_{0}}, b$, and then use
\eqref{a.27} to obtain the following background
\begin{eqnarray}
d {s}^2 &=& -a_{_{0}}^2 {d t}^2   -a_{_{0}}^2  e^{-2t} d x_1~ d x_2 + b ~{d x_3}^2,\label{x.15}\\
{B} &=& a_{_{0}}^2~x_1~ e^{-2t} ~d x_2 \wedge d t.\label{x.16}
\end{eqnarray}
Comparing \eqref{x.15} and  the general form of the string cosmology metric
\begin{eqnarray}
d {s}^2 = -g_{_{00}}^2(t) {d t}^2  +\sum_{a,b=1}^{3} {R_{\mu}^{~a}~R_{\nu}^{~b}~ g_{_{ab}}(t)~dx^{\mu}~dx^{\nu}},~~\label{x.17}
\end{eqnarray}
one concludes that \eqref{x.15} is nothing but the Bianchi type III string cosmology  metric.
This metric  has $(2 , 2)$-signature if $b$ is considered positive.
One can easily check that the metric \eqref{x.15} and the field strength corresponding to the $B$-field \eqref{x.16}
($H_{_{x_1 x_2 t}}={a_{_{0}}^2}e^{-2t}/{2}$) along with the dilaton field
$\phi=\upsilon_{_{0}} + \upsilon_{_{1}} x_3$ (for some constants $\upsilon_{_{0}}, \upsilon_{_{1}}$) make up a solution for the vanishing of the one-loop $B$-functions equations \eqref{d.6.1.28}-\eqref{d.6.1.30}.
It is also interesting to note that the corresponding action to
\eqref{x.15} and \eqref{x.16} is equivalent
to the $GL(2,\mathbb{R})$ WZW model.
This means that the obtained background  can be described as
an exact CFT.

The dual model is constructed  on $3+1$-dimensional
manifold ${\tilde M} \approx O \times {\tilde {\bf G}}$ with ${\tilde {\bf G}} =3A_1$. Finally, using
\eqref{x.13} and \eqref{x.14} together with equations \eqref{a.32}-\eqref{a.35}
the dual background is obtained to be
\begin{eqnarray}
{d {\tilde s}}^2 &=&-a_{_{0}}^2 {d t}^2 + \frac{1}{b}~ {d x_{_{3}}}^2\nonumber\\
&&+ \frac{a_{_{0}}^2 e^{-2t}}{ {{\tilde x}_2}^2-
\frac{a_{_{0}}^4}{4}e^{-4t}}\big(d {{\tilde x}_1}~d {{\tilde x}_2}
+a_{_{0}}^2 e^{-2t}~ d {{\tilde x}_1}~d t\big),\label{x.18}\\
{\tilde B} &=&  \frac{{{\tilde x}_2}}{ {{\tilde x}_2}^2-
\frac{a_{_{0}}^4}{4}e^{-4t}}\big(d {{\tilde x}_1} \wedge d {{\tilde x}_2}
+a_{_{0}}^2 e^{-2t} d {{\tilde x}_1}  \wedge d t\big).~~~~~~~\label{x.19}
\end{eqnarray}
The dilaton field  that supports the dual background is obtained in the following form
\begin{eqnarray}
{\tilde \phi} ~=~ \vartheta_{_{0}} + \vartheta_{_{1}} {\tilde x}_3
-\log \Big(\frac{a_{_{0}}^2+2{\tilde x}_2  e^{2t}}{a_{_{0}}^2-2{\tilde x}_2  e^{2t}}\Big),\label{x.20}
\end{eqnarray}
where $\vartheta_{_{0}}, \vartheta_{_{1}}$ are some constants.

\section{\label{Sec.IV}Conformal invariance of the T-dual models up to two-loop order (first
order in  $\alpha'$)}

So far, we have been concerned with the conformal invariance of the T-dual models up to one-loop order  (zeroth order in $\alpha'$).
As mentioned in Sec. \ref{Sec.III}, the conditions for conformal
invariance of the $\sigma$-model with action \eqref{a.3} can be interpreted as field equations for
${G}_{_{\mu \nu}}$, ${B}_{_{\mu \nu}}$ and $\phi$ of the string effective action \cite{{A.Sen},{callan}}. These equations to the first
order in  $\alpha'$  take the following form \cite{c.hull}
\begin{widetext}
\begin{subequations}
\begin{eqnarray}
&&{\cal R}_{{\mu \nu}}-(H^2)_{{\mu \nu}}+{\nabla}_\mu
{\nabla}_\nu \phi +\frac{1}{2} \alpha' \Big[{\cal R}_{{\mu \rho \sigma \lambda}} {\cal R}_{\nu}^{{~\rho\sigma \lambda}}
+2 {\cal R}_{{\mu \rho\sigma \nu }} (H^2)^{\rho\sigma}+2 {\cal R}_{{\rho\sigma \lambda(\mu }}H_{\nu)}^{~\lambda \delta} H^{\rho\sigma}_{~~_{\delta}} +\frac{1}{3} ({\nabla}_\mu H_{\rho\sigma\lambda})
({\nabla}_\nu H^{\rho\sigma\lambda})\nonumber\\
&&~~~~~~~~~~~~~~~~~~~~~~-({\nabla}_\lambda H_{\rho\sigma \mu})
({\nabla}^\lambda H^{\rho \sigma}_{~~\nu})+2 H_{{\mu \rho \sigma}} H_{{\nu \lambda \delta}} H^{{\eta \delta \sigma}} H_{\eta}^{~~\lambda \rho}
+2 H_{{\mu  \rho \sigma }} H_{{\nu \lambda}} ^{~~\sigma}   (H^2)^{\lambda \rho} \Big]+
{\cal O}(\alpha'^2)~=~0,\label{e.1}\\
&&{\nabla}^\lambda H_{{\lambda \mu \nu}} -  ({\nabla}^\lambda\phi')  H_{{\mu \nu \lambda}}
+\alpha' \Big[{\nabla}^\lambda H^{\rho \sigma}_{~~[\mu}{\cal R}_{_{\nu]} \lambda \rho \sigma} -({\nabla}_\lambda H_{\rho\mu\nu}) (H^2)^{\lambda\rho}-2 ({\nabla}^\lambda H^{\rho \sigma}_{~~[\mu})H_{_{\nu]} \rho \delta } H_{\lambda \sigma}^{~\;\delta}\Big]
+{\cal O}(\alpha'^2)~=~0,~~~~~~~\label{e.2}\\
&&2 \Lambda + {\nabla}^2 \phi' - ({\nabla} \phi')^2+\frac{2}{3} H^{{2}}
-\alpha' \Big[\frac{1}{4} {\cal R}_{{\mu \rho \sigma \lambda}} {\cal R}^{{\mu \rho \sigma \lambda}}
-\frac{1}{3} ({\nabla}_\lambda H_{\mu \nu \rho })
 ({\nabla}^\lambda H^{\mu \nu \rho })
-\frac{1}{2} H^{\mu\nu}_{~~\lambda} H^{\rho \sigma \lambda} {\cal R}_{\mu \nu \rho \sigma }\nonumber\\
&&~~~~~~~~~~~~~~~~~~~~~~~~~~~~~~~~ -{\cal R}_{\mu \nu} (H^2)^{\mu \nu} +\frac{3}{2} (H^2)_{\mu \nu} (H^2)^{\mu \nu}
+\frac{5}{6} H_{\mu \nu \rho } H^{\mu}_{~~\sigma \lambda} H^{\nu \sigma}_{~~\delta} H^{\rho \lambda \delta}\Big] +{\cal O}(\alpha'^2)~=~0
,\label{e.3}
\end{eqnarray}
\end{subequations}
\end{widetext}
where $\phi' = \phi + \alpha' q H^2$ for some coefficient $q$ \cite{c.hull}, $(H^2)^{\mu \nu} = H^{\mu \rho \sigma } H_{\rho \sigma}^{~~\nu}$ and
${\cal R}_{{\mu \rho \sigma \lambda}}$ is the Riemann tensor field.  We note that round brackets
denote the symmetric part on the indicated indices whereas square brackets
denote the antisymmetric part.
Below using the above equations  we check the conformal invariance conditions of the T-dual models  up to two-loop order (first
order in  $\alpha'$). In fact, we introduce new solutions for two-loop  $B$-function equations of the  $\sigma$-model with a non-vanishing field strength
$H$ and the dilaton field in both cases of the
absence and presence of a cosmological constant $\Lambda$.

$\bullet$~ As shown in  subsection A of Sec. \ref{Sec.III}, the background of the original $\sigma$-model \eqref{d.5} is given by the formulas \eqref{d.6.1}
and \eqref{d.6.2} so that this model is  equivalent to  the $H_4$ WZW model. Therefore, it should be conformally invariant.
In the case of this model, the only non-vanishing component of the Riemann tensor is ${\cal R}_{_{x_1 y_1 x_2 y_1}}=-(e^{x_1+y_1})/4$.
Moreover, the only non-vanishing component of $(H^2)^{^{\mu\nu}}$ is $(H^2)^{^{y_2y_2}}=-1/2$ and all components of
${\nabla}_\lambda H_{{\mu \nu \sigma}}$  vanish. Hence using these results, the field equations \eqref{e.1}-\eqref{e.3} are satisfied for the metric \eqref{d.6.1} and the tensor field \eqref{d.6.2} together with
the dilaton field \eqref{d.6.1.1} and zero cosmological constant as this was expected.

$\bullet$~ In order to investigate the conformal invariance conditions of the dual model to the $H_4$ WZW (the $\sigma$-model \eqref{d.11})
up to the first order in  $\alpha'$, we first find that the only non-vanishing components of $({\tilde H}^2)_{_{\mu\nu}}$
and $({\tilde H}^2)^{^{\mu\nu}}$ are $({\tilde H}^2)_{_{y_{_{1}}y_{_{1}}}}=
-(e^{y_{_{1}}} + {\tilde x}_2)^2/2(e^{y_{_{1}}} - {\tilde x}_2)^2$ and
$({\tilde H}^2)^{^{y_{_{2}} y_{_{2}}}}=({\tilde H}^2)_{_{y_{_{1}}y_{_{1}}}}$, respectively.
Also, the only non-vanishing components of
${\tilde {\nabla}}_{_{\lambda}} {{\tilde H}}_{_{\mu \nu \rho}}$ may be expressed as
\begin{eqnarray}
{\tilde {\nabla}}_{_{{\tilde x}_2}} {{\tilde H}}_{_{{\tilde x}_1 {\tilde x}_2 {y_{_{1}}}}} &=&
\frac{e^{2{y_{_{1}}}}}{(e^{y_{_{1}}} + {\tilde x}_2)(e^{y_{_{1}}} - {\tilde x}_2)^3},\nonumber\\
{\tilde {\nabla}}_{_{{y_{_{1}}}}} {{\tilde H}}_{_{{\tilde x}_1 {\tilde x}_2 {y_{_{1}}}}} &=&-
\frac{{\tilde x}_2 e^{2 {y_{_{1}}}}}{(e^{y_{_{1}}} + {\tilde x}_2)(e^{y_{_{1}}} - {\tilde x}_2)^3}.\label{e.4}
\end{eqnarray}
Using these  results together with the given data for this model in subsection A of Sec. \ref{Sec.III}, one verifies the field equations \eqref{e.1} and \eqref{e.2}
for the metric \eqref{d.12} and the tensor field \eqref{d.13} along with
the dilaton field \eqref{d.14}. The equation \eqref{e.3} is also satisfied with ${\tilde \Lambda}=0$.

$\bullet$~ Under the coordinate transformation \eqref{d.21}, the background of the original $\sigma$-model (\ref{d.16}) was represented by
\eqref{d.22} and \eqref{d.23}. It was shown that resulting  background as an exact CFT satisfies the
vanishing of the one-loop $B$-functions equations \eqref{d.6.1.28}-\eqref{d.6.1.30} with the dilaton field \eqref{d.24}.
Using the expressions \eqref{d.22} and \eqref{d.23} for the background fields
one may verify that the only non-vanishing components of Riemann tensor are
${\cal R}_{_{t \varphi t \varphi}} ={r^4}/{l^2},~
{\cal R}_{_{t r t r}} =1/{l^2}$ and ${\cal R}_{_{\varphi r \varphi r}}=-1$; consequently, the Kretschmann scalar is computed to be $K=12$.
Moreover, we get that the only non-vanishing components of $(H^2)^{^{\mu\nu}}$ are
$(H^2)^{^{tt}}=(2l^2)/r^2,~ (H^2)^{^{\varphi\varphi}}=-2/r^2$ and $(H^2)^{^{rr}}=-2r^2$, and all components of
${\nabla}_\lambda H_{{\mu \nu \sigma}}$ vanish. Putting these pieces together, one verifies
equations \eqref{e.1} and \eqref{e.2} with the dilaton field \eqref{d.24}.
It is then interesting to note that in this case the field equation \eqref{e.3} is satisfied
if the following relation is held between the constants ${\zeta_{_{1}}}^2 $, $\Lambda$, $b$ and  $\alpha'$:
\begin{eqnarray}
\alpha'~=~-\frac{1}{4}(2+\frac{{\zeta_{_{1}}}^2 }{2b}-\Lambda).\label{e.5}
\end{eqnarray}

$\bullet$~ As mentioned in the preceding section, the dual model of the $GL(2,\mathbb{R})$ WZW (equations
\eqref{x.6}-\eqref{x.8}) does satisfy the vanishing of the one-loop $B$-functions equations. Unfortunately,
this background
does not satisfy the equations for the two-loop $B$-functions. One can show that for this background
all equations \eqref{e.1}-\eqref{e.3} are satisfied except for the components of
$B^{^{G}}_{ii}, (i=t, x, r)$, $B^{^{B}}_{tx}$ and $B^{^{\Phi}}$.

\section{\label{Sec.V}Summary and concluding remarks}

Using the PL T-duality approach in the presence of spectators we have constructed some
non-Abelian T-dualizable $\sigma$-models  on $2+2$-dimensional
target manifolds $M \approx O \times \bf G$ and
${\tilde M} \approx O \times {\bf {\tilde G}}$, where $\bf G$ and ${\bf {\tilde G}}$ are two-dimensional
real non-Abelian and Abelian Lie groups, respectively.
We have shown that the original $\sigma$- models are  equivalent to the $H_4$ and $GL(2,\mathbb{R})$
WZW models.
In this way, we could obtain some new T-dual backgrounds for these  WZW models.
The most interesting feature of our results is the invariance  of  the $H_4$ WZW model
under the non-Abelian T-duality.
We have shown that the $GL(2,\mathbb{R})$ WZW model as a T-dualizable $\sigma$- model is  equivalent to
$AdS_{3} \times \mathbb{R}$ space and has  no horizon
and no curvature singularity, while the dual spacetime of the $GL(2,\mathbb{R})$ WZW model is stationary and asymptotically
flat and has a single horizon and a curvature singularity. Moreover,
it was shown that for the line element \eqref{d.22}, the Killing vectors $\partial/{\partial t}$ and
$(1/b)~\partial/{\partial z}$ with $b<0$  are
time-like. Analogously, one can show that the dual line element \eqref{x.6}  possesses
three independent Killing vectors $\sqrt{3} \partial/{\partial t}$, ${3}/{2}(\sqrt{3} \partial/{\partial t} - \partial/{\partial x})$ and $(1/b)~\partial/{\partial z}$. The first two Killing vectors become
time-like for the ranges $r>2$ and $r>4/3$, respectively. The last Killing vector stays
everywhere time-like for $b<0$.
Hence, the duality has involved the time-like directions.
In summary, in the case of
the effect of the non-Abelian T-duality (here as the PL T-duality on a semi-Abelian double)
on the $GL(2,\mathbb{R})$ WZW model three points have been highlighted.\\
${1.}$ The non-Abelian T-duality
transformation  has changed the asymptotic behavior of solutions from $AdS_{3} \times \mathbb{R}$ to flat space.\\
${2.}$ This transformation has related a solution with no horizon
and no curvature singularity  to a solution with a single horizon and a curvature singularity.\\
${3.}$ The duality has involved the time-like directions.

We have also obtained the non-critical Bianchi type III string cosmological model with a
non-vanishing field strength from a T-dualizable $\sigma$-model
and have shown that this model describes an exact CFT.
Most importantly, we have discussed the conformal invariance of the T-dual $\sigma$- models such that the duals
of the $H_4$ WZW model are conformally invariant up to the first order in $\alpha'$, while
the conformal invariance condition for the dual spacetime of the $GL(2,\mathbb{R})$ WZW model
has only been satisfied up to zeroth order in $\alpha'$.

As we have shown, all our models  satisfy the vanishing of the  one-loop Beta-functions equations.
Therefore, each pair of them consists of two canonically equivalent models.
Among these models, only \eqref{d.5} and \eqref{d.14.2} and their dual pairs \eqref{d.11} and \eqref{d.14.4}, respectively, satisfy
the equations for two-loop $B$-functions.

The findings of our study showed that $2+2$-dimensional  manifold $M \approx O \times \bf G$ with two-dimensional
real non-Abelian Lie group ${\bf G}=A_2$ is weathly. In addition to  PL symmetric backgrounds constructed out in this
paper one can obtain  other string and gravitational backgrounds from mutually T-dualizable $\sigma$- models on manifold $M \approx O \times \bf G$
with ${\bf G }= A_2$ when the dual manifold is ${\tilde M } \approx O \times {\bf {\tilde G}}$
with $\tilde {\bf G }= 2A_1$ .
In this regard, the following further developments come to mind.

\smallskip

$\bullet$~ {\bf Plane-parallel (pp-)wave:} Homogenous plane wave is generally defined by the metric of the following form \cite{pp-wave1}
\begin{eqnarray}
ds^2~=~2 du dv - A_{_{\mu \nu}}(u)~ X^{\mu} X^{\nu} du^2+ dX^2,\label{conclusion.1}
\end{eqnarray}
where $dX^2$ is the standard metric on Euclidean space $E^d$ and $X\in E^d$.
A special case of isotropic homogenous plane wave metric can be chosen by
$A_{_{\mu \nu}}(u) = \lambda(u) {\delta}_{_{\mu \nu}}$. Furthermore,
for special choice of $\lambda(u)=k/u^2$, the metric becomes \cite{pp-wave2}
\begin{eqnarray}
ds^2~=~2 du dv - \frac{k}{u^2}~ (x^2+y^2) du^2+ dx^2+dy^2,\label{conclusion.2}
\end{eqnarray}
where $k$ is an arbitrary real constant. The metric \eqref{conclusion.2} does satisfy
the conformal invariance conditions equations up to the first
order in  $\alpha'$, equations  \eqref{e.1}-\eqref{e.3}, with zero field strength. The cosmological constant $\Lambda$
in this case vanishes and dilaton field is obtained to be \cite{pp-wave2}
\begin{eqnarray}
\phi &=&\gamma_{_{0}}+ \gamma_{_{1}} u + 2k \log u,\label{conclusion.3}
\end{eqnarray}
where $\gamma_{_{0}}$ and $\gamma_{_{1}}$  are  the constants of integration. Since
the field strength $H$ is zero, one can easily consider explicit expressions for the field $B$ in such a way that
the terms concerning $B$-field in action of $\sigma$-model
contribute to the Lagrangian as the total derivatives, which can be ignored.
To obtain the non-Abelian T-dual geometry of pp-wave background in the PL T-duality approach
with spectators,  we first construct
the original $\sigma$-model corresponding to the pp-wave metric \eqref{conclusion.2}.
In this case, a convenient choice of the spectator-dependent matrices may be expressed as
{\small{\begin{eqnarray}\label{conclusion.5}
E^{+}_{0\;ab}&=&\left( \begin{array}{cc}
                    -k({y_{_{1}}}^2+{y_{_{2}}}^2) & 1\\
                     1 & 0
                      \end{array} \right),~~
F_{_{\alpha\beta}} =\left( \begin{array}{cc}
                    1 & 0\\
                    0 & 1
                      \end{array} \right),\nonumber\\
                       F^{+^{(1)}}_{a \beta}&=&0,~~~~~~~~~F^{+^{(2)}}_{\alpha b}=0.
\end{eqnarray}}}
Inserting \eqref{conclusion.5} into equations \eqref{a.13}-\eqref{a.16} and  noting that $\Pi^{ab}(g)$  is zero, the action of original $\sigma$-model \eqref{a.27} yields
\begin{eqnarray}
S &=& \frac{1}{2} \int d \sigma^+ d \sigma^-~\Big[\partial_+
 y_{_{1}} \partial_- y_{_{1}}+ \partial_+ y_{_{2}} \partial_- y_{_{2}}\nonumber\\
&&~~~~~~~~~~+e^{x_{_{1}}}(\partial_+ x_{_{1}} \partial_- x_{_{2}}+
\partial_+ x_{_{2}} \partial_- x_{_{1}})\nonumber\\
&&~~~~~~~~~~-k({y_{_{1}}}^2+{y_{_{2}}}^2)\partial_+ x_{_{1}} \partial_- x_{_{1}}
\Big].\label{conclusion.6}
\end{eqnarray}
Carrying out the coordinates transformation
${x_1} \rightarrow  \ln u, x_2 \rightarrow  v,  y_{_{1}} \rightarrow  x ,  y_{_{2}} \rightarrow  y$ one arrives at
the pp-wave metric \eqref{conclusion.2} from action \eqref{conclusion.6}.
Thus, inserting \eqref{d.9} and \eqref{conclusion.5} into equations \eqref{a.32}-\eqref{a.35} and then using \eqref{a.31}
one can obtain a non-Abelian T-dual $\sigma$-model to \eqref{conclusion.6}.

\smallskip

$\bullet$~ {\bf  G\"{o}del and G\"{o}del-type metrics:}
Among the known exact solutions of Einstein field equations  gravity,
the G\"{o}del and G\"{o}del-type metrics \cite{godel} play a special role. It was shown within the usual general relativity
that these solutions describe rotating universes, and allow for the existence of closed time-like curves. These metrics are compatible with incoherent matter distribution at rest and can be described by the line element looking like
\begin{eqnarray}
ds^2&=&l^2\big[-dt^2 +(\beta-1) r^2 d\varphi^2\nonumber\\
&&~~~~~~~~~~-2r  dt d \varphi + \frac{dr^2}{r^2} +dz^2\big],\label{conclusion.15}
\end{eqnarray}
for some constants $l$, $\beta$. The metric is a direct product of $\mathbb{R}$ associated with the
coordinate $z$ and the three-dimensional metric of $(t, \varphi, r)$. The
original G\"{o}del metric \cite{godel} is recovered when we take $\beta=1/2$.
In Ref. \cite{Barrow} it has been shown that the G\"{o}del metric can be considered as exact solutions
in string theory for the full ${\cal O}(\alpha')$ action including both dilaton field $\phi$ and field strength $H$.
Following Ref. \cite{Barrow} we assume that the dilaton field
depends only on the $z$ coordinate, so  $\phi(z) = \phi_{_{0}}+fz$ for some constants $\phi_{_{0}}$, $f$.
With this assumption and  taking the zero field strength, $H=0$,
the field equations \eqref{e.1} and \eqref{e.2} are satisfied for the  metric \eqref{conclusion.15} in such a way that
the inverse string tension $\alpha'$ has to satisfy relation
$\alpha' = 4l^2 \beta$  only with $\beta=1$ or $\beta=3/4$.
Finally, the field equation \eqref{e.3} is satisfied
if the following relation is held between the constants $f$, $\Lambda$ and  $l$:
\begin{eqnarray}
f^2~=~ \left\{\begin{array}{ll} -\frac{3}{4} +2 \Lambda l^2 &{\rm if} \;{\beta} =1\nonumber\\
-\frac{2}{3} +2 \Lambda l^2 &{\rm if} \;{\beta} =\frac{3}{4}\end{array} \right..
\end{eqnarray}
In addition, one can check that the metric \eqref{conclusion.15} for ${\beta} =1/2$  (the original G\"{o}del metric)
along with the respective $B$-field and dilaton field
\begin{eqnarray}
B&=&\frac{l^2}{2} r ~ d\varphi \wedge dt,\nonumber\\
\phi &=&\phi_{_{0}}+f z,\nonumber
\end{eqnarray}
satisfy the field equations \eqref{e.1}-\eqref{e.3} provided that $\alpha' = -l^2/2$ and $f^2 = 1/2+ 2 \Lambda l^2$.
Now, by using the above results and by choosing the appropriate spectator-depended background  matrices we can construct a $\sigma $-model including
the G\"{o}del (G\"{o}del-type) metric in the form \eqref{conclusion.15} and the
given $B$-fields. In this way, one can study the non-Abelian T-duality of the G\"{o}del (G\"{o}del-type) metric.
We intend to address this problem in the future.

\smallskip
$\bullet$~ {\bf $\bf AdS_4$ metric:} $AdS_4$  metric with radius $l$ and  constant negative scalar curvature $\Lambda =-12/l^2$  is one of the maximal symmetric four-dimensional spacetimes. A simple form of this metric in coordinates $(z, x^+, x^-, \rho)$  is given by
\begin{eqnarray}
ds^2~=~\frac{l^2}{z^2} \big(dz^2 - dx^+  d x^- +d\rho^2\big).\label{conclusion.16}
\end{eqnarray}
Solving  the field equations \eqref{e.1}-\eqref{e.3} for metric \eqref{conclusion.16}  one should be so lucky
to obtain an appropriate field strength along with a dilaton field.
Then he/she can study the non-Abelian T-duality of the $AdS_4$  metric.

\acknowledgments{The author gratefully thanks to the Referees for the constructive
comments and recommendations
which definitely help to improve the readability and quality of the paper.}


\appendix

\section{The WZW models based on the $H_4$ and $GL(2 , \mathbb{R})$ Lie groups}

In this appendix, we  construct the WZW models based on the $H_4$ and $GL(2 , \mathbb{R})$ Lie groups.
To define a WZW model, in general, given a Lie algebra with generators $T_a$ and structure constants $f^c_{~ab}$, one needs a non-degenerate ad-invariant symmetric bilinear form $\Omega_{ab} = <T_a~ , ~T_b >$  on Lie algebra ${\cal G}$
so that it satisfies the following relation \cite{Witten}
\begin{eqnarray}\label{A.1}
f^d_{~ab} \;\Omega_{dc}+ f^d_{~ac} \;\Omega_{db}\;=\;0.
\end{eqnarray}
The WZW model based on a Lie group $\bf G$  is defined on a Riemannian surface $\Sigma$
as a worldsheet by the following action \cite{Witten}
\begin{eqnarray}\label{A.2}
I(g) &=&  \frac{1}{2} \int_{\Sigma} d\sigma^+ d\sigma^-\;
{\Omega}_{ab}~L^{\hspace{-0.5mm}a}_{+}\;L^{\hspace{-0.5mm}b}_{-}\nonumber\\
&&+\frac{1}{12} \int_{B} d^3 \sigma~
\varepsilon^{ \gamma \alpha \beta}
L^{\hspace{-0.5mm}a}_{\gamma}
\;L^{\hspace{-0.5mm}b}_{\alpha}~  L^{\hspace{-0.5mm}c}_{\beta}~  {\Omega}_{ad}\;
\;f^{d}_{~bc},~~~~~~
\end{eqnarray}
where {\small $B$} is a three-manifold bounded by worldsheet $\Sigma$, and the components of the left invariant one-forms
$L^{\hspace{-0.5mm}a}_{\alpha}$'s are defined via $g^{-1}
\partial_{\alpha}g\;=\;L^{\hspace{-0.5mm}a}_{\alpha}~ T_{a}$ in which $g:  \Sigma \rightarrow  \bf G$ is an element of Lie group $\bf G$.

\subsection{\label{app:subsec.1}The $H_4$ WZW model}

Before proceeding to construct the model, let us first introduce  the oscillator Lie algebra $h_4$ of the  Lie group $H_4$.
The  Lie algebra $h_4$ is generated by the generators $\{N, A_+, A_-, M\}$ with the following non-zero Lie brackets
\begin{eqnarray}\label{A.3}
[N , A_+]=A_+,~~[N , A_-]=-A_-,~~[A_- , A_+]=M.~~
\end{eqnarray}
Using (\ref{A.1}) and (\ref{A.3}) one can simply get a non-degenerate ad-invariant bilinear form $\Omega_{ab}$ on $h_4$ as
{\small \begin{eqnarray}\label{A.4}
\Omega_{ab}~=~\left( \begin{tabular}{cccc}
              $0$ & $0$ & $0$ & $-1$\\
              $0$ & $0$ & $1$ & $0$ \\
              $0$ & $1$ & $0$ & $0$ \\
              $-1$ & $0$ & $0$ & $0$\\
                \end{tabular} \right).
\end{eqnarray}}
In order to calculate the $L^{\hspace{-0.5mm}a}_{\alpha}$'s on the  Lie group $H_4$ we parameterise an element of  $H_4$  as
\begin{eqnarray}\label{A.5}
g\;=\; e^{m M} ~ e^{a_- A_-}~ e^{n N} ~ e^{a_+ A_+},
\end{eqnarray}
Finally, the WZW action on the $H_4$ Lie group is worked out to be of the form \cite{eghbali11}
\begin{eqnarray}
I(g) &=& \frac{1}{2} \int d \sigma^+ d \sigma^-~\Big[-\partial_+ n \partial_- m-\partial_+ m \partial_- n\nonumber\\
&&~~~~+ e^{n} (\partial_+ a_+ \partial_- a_-
+\partial_+ a_- \partial_- a_+)\nonumber\\
&&~~~~+a_+ e^{n}(\partial_+ a_- \partial_- n - \partial_+ n \partial_- a_-)\Big].\label{A.6}
\end{eqnarray}

\subsection{\label{app:subsec.2}The $GL(2,\mathbb{R})$ WZW model}

The $gl(2,\mathbb{R})$ Lie algebra is  spanned by the generators $\{J_3, J_+, J_-,  I\}$
which obey the following commutation rules
\begin{eqnarray}\label{A.7}
{[J_3 , J_+]}&=&2J_+,~~~~[J_3 , J_-]=-2J_-,\nonumber\\
{[J_+ , J_-]}&=&J_3,~~~~~~~~[I~ ,~ .]=0.
\end{eqnarray}
where $I$ is the central generator. We notice that $gl(2,\mathbb{R}) = sl(2,\mathbb{R}) \oplus u(1)$.
Using (\ref{A.7}), a non-degenerate solution to  (\ref{A.1}) is obtained to be of the form
{\small \begin{eqnarray}\label{A.8}
\Omega_{ab}~=~\left( \begin{tabular}{cccc}
              $2a$ & $0$ & $0$ & $0$\\
              $0$ & $0$ & $a$ & $0$ \\
              $0$ & $a$ & $0$ & $0$ \\
              $0$ & $0$ & $0$ & $b$\\
                \end{tabular} \right),
\end{eqnarray}}
for some non-zero constants  $a$, $b$.
In order to construct the WZW model based on the $GL(2,\mathbb{R})$ Lie group
we parameterize the $GL(2,\mathbb{R})$ with coordinates $\{\theta_3, \theta_+, \theta_-,  \theta\}$
so that its elements can be written as
\begin{eqnarray}\label{A.9}
g\;=\; e^{\theta_+ J_+} ~ e^{\theta_3 J_3} ~ e^{\theta_- J_-}~ e^{\theta I }.
\end{eqnarray}
Using (\ref{A.9}), we then obtain
\begin{eqnarray}\label{A.10}
L^{\hspace{-0.5mm}J_{_{3}}}_{\pm}&=&\theta_{_{-}} e^{-2 \theta_{_{3}}} \partial_{_{\pm}} \theta_{_{+}} +\partial_{_{\pm}} \theta_{_{3}},\nonumber\\
L^{\hspace{-0.5mm}J_{_{+}}}_{\pm}&=&e^{-2 \theta_{_{3}}} \partial_{_{\pm}} \theta_{_{+}},\nonumber\\
L^{\hspace{-0.5mm}J_{_{-}}}_{\pm}&=& - \theta_{_{-}}^2 e^{-2 \theta_{_{3}}} \partial_{_{\pm}} \theta_{_{+}} -2 \theta_{_{-}}
 \partial_{_{\pm}} \theta_{_{3}} +  \partial_{_{\pm}} \theta_{_{-}},\nonumber\\
L^{\hspace{-0.5mm}I}_{\pm}&=& \partial_{_{\pm}} \theta.
\end{eqnarray}
Finally, the $GL(2,\mathbb{R})$  WZW action looks like
\begin{eqnarray}
I(g) &=& \frac{1}{2} \int d \sigma^+ d \sigma^-~\Big[b~ \partial_{_{+}} \theta \partial_{_{-}} \theta
\nonumber\\
&&+2a\big(\partial_{_{+}} \theta_{_{3}} \partial_{_{-}} \theta_{_{3}}+ e^{-2 \theta_{_{3}}} \partial_{_{+}}
\theta_{_{-}} \partial_{_{-}} \theta_{_{+}}\big)\Big].~~~\label{A.11}
\end{eqnarray}



\end{document}